\documentclass[11pt]{article}

\usepackage[margin=1in]{geometry}
\usepackage{times}

\usepackage{tikz}
\usepackage{graphicx}
\graphicspath{{figures/}{figures/pcf_subset_eval/}{figures/pcf_insights/}{../output/figures/}{../output/figures/pcf_subset_eval/}{../output/figures/pcf_insights/}}
\usepackage{inconsolata}
\usepackage{xcolor}
\usepackage{float}
\usetikzlibrary{shapes.geometric, shapes.symbols, arrows.meta, positioning, calc, backgrounds}
\usepackage{wrapfig}
\usepackage{tcolorbox}
\tcbuselibrary{skins}
\usepackage{seqsplit}

\definecolor{tealfill}    {HTML}{E1F5EE}
\definecolor{tealstroke}  {HTML}{0F6E56}
\definecolor{tealtext}    {HTML}{085041}
\definecolor{purplefill}  {HTML}{EEEDFE}
\definecolor{purplestroke}{HTML}{534AB7}
\definecolor{purpletext}  {HTML}{3C3489}
\definecolor{amberfill}   {HTML}{FAEEDA}
\definecolor{amberstroke} {HTML}{854F0B}
\definecolor{ambertext}   {HTML}{633806}
\definecolor{bluefill}    {HTML}{E6F1FB}
\definecolor{bluestroke}  {HTML}{185FA5}
\definecolor{bluetext}    {HTML}{0C447C}
\definecolor{coralfill}   {HTML}{FAECE7}
\definecolor{coralstroke} {HTML}{993C1D}
\definecolor{coraltext}   {HTML}{712B13}
\definecolor{greenfill}   {HTML}{EAF3DE}
\definecolor{greenstroke} {HTML}{3B6D11}
\definecolor{greentext}   {HTML}{27500A}
\definecolor{grayfill}    {HTML}{F1EFE8}
\definecolor{graystroke}  {HTML}{5F5E5A}
\definecolor{graytext}    {HTML}{2C2C2A}

\tikzset{
  procnode/.style={rectangle, rounded corners=6pt, minimum width=3.6cm,
    minimum height=1.0cm, align=center, text width=3.4cm,
    draw, line width=0.5pt, font=\small},
  dbnode/.style={cylinder, shape border rotate=90, minimum width=2.8cm,
    minimum height=1.1cm, align=center, text width=2.4cm,
    draw, line width=0.5pt, font=\small, aspect=0.25},
  teal/.style  ={fill=tealfill,   draw=tealstroke,   text=tealtext},
  purple/.style={fill=purplefill, draw=purplestroke, text=purpletext},
  amber/.style ={fill=amberfill,  draw=amberstroke,  text=ambertext},
  blue/.style  ={fill=bluefill,   draw=bluestroke,   text=bluetext},
  coral/.style ={fill=coralfill,  draw=coralstroke,  text=coraltext},
  green/.style ={fill=greenfill,  draw=greenstroke,  text=greentext},
  gray/.style  ={fill=grayfill,   draw=graystroke,   text=graytext},
  solidarr/.style={-{Latex[length=5pt,width=4pt]}, line width=0.8pt, color=graystroke},
  crossarr/.style={-{Latex[length=5pt,width=4pt]}, line width=1.0pt, color=tealstroke}
}
\newcommand{\sub}[1]{\\[2pt]{\footnotesize\color{graystroke}#1}}

\usepackage{url}
\usepackage[numbers]{natbib}

\usepackage{amsmath}
\usepackage{amssymb}

\usepackage{graphicx}
\usepackage{float}
\usepackage{multirow}
\usepackage{subcaption}
\graphicspath{{figures/}}

\usepackage[toc,page]{appendix}

\usepackage{xcolor} 
\usepackage{tikz}
\usetikzlibrary{positioning, arrows.meta, calc, shapes, fit, matrix}

\definecolor{nncolor}{RGB}{200,200,200}       
\definecolor{randcolor}{RGB}{180,210,255}     
\definecolor{currcolor}{RGB}{255,200,180}     
\definecolor{boxbg}{RGB}{245,245,245}

\title{\textbf{Trading Engagement for Sustainability: Carbon-Aware Re-ranking for E-commerce Recommendations}}

\author{
Anders Vestrum\thanks{Equal contribution.} \\
University of California, Berkeley
\and
Jørgen Bergh\footnotemark[1] \\
University of California, Berkeley
\and
Noah Lund Syrdal\footnotemark[1] \\
University of California, Berkeley
}

\date{%
\today
}

\begin{document}

\maketitle

\begin{abstract}

E-commerce recommender systems strongly influence which products users consider and purchase, yet sustainability signals such as Product Carbon Footprint (PCF) are almost never available at catalog scale. We study carbon-aware product recommendation in the realistic setting where PCF labels are missing for most items and must be inferred. We first estimate product-level carbon footprints via a retrieval-augmented PCF estimation pipeline that transfers supervision from the Carbon Catalogue (a small set of life-cycle-assessed products) to a large unlabeled e-commerce catalog using semantic similarity search, few-shot LLM prompting, and a nearest-neighbour fallback. We then apply a carbon-aware post-hoc re-ranking strategy on top of relevance scores produced by three established recommendation models (BPR, NeuMF, LightGCN), trading off predicted user--item engagement against estimated carbon footprint through a single tunable parameter $\lambda$. In this offline study, engagement is operationalized through Amazon review interactions, which serve as implicit feedback and as a proxy for user interest or purchase behavior. We evaluate the framework on the Amazon Reviews dataset across three product categories: \textit{Home and Kitchen}, \textit{Sports and Outdoors}, and \textit{Electronics}. By sweeping $\lambda$, we construct Pareto frontiers that characterize the achievable engagement and carbon trade-off for each model and category. Substantial carbon reductions are achievable at minimal engagement cost across
all models and categories. However, the available carbon headroom varies by model
and category, underscoring the importance of model choice and domain context. \footnote{Code available at: \url{https://github.com/andersvestrum/carbon-aware-recsys}.}

\end{abstract}

\section{Introduction}
\label{sec:introduction}

Online retail has grown into one of the dominant channels through which consumers acquire
goods, and the environmental consequences of that growth have attracted sustained attention
across dimensions including shipping logistics, packaging waste, and return-driven reverse
supply chains~\cite{olah2023sustainable}. The scale of the problem is considerable: UN
Trade and Development estimates that global e-commerce sales reached USD~27 trillion in
2022~\cite{unctad2024ecommerce}, and the carbon consequences of that volume are
increasingly documented as a meaningful contributor to greenhouse gas
emissions~\cite{buldeorai2023ecommerce}. The environmental footprint of online retail is
not uniform. It depends in substantial part on how purchasing decisions are structured and
presented to consumers~\cite{buldeorai2023ecommerce}.

At the policy level, this urgency has found expression in legislation. The European Union's
Directive 2024/825 on Empowering Consumers for the Green Transition strengthens rules around
environmental claims and consumer-facing sustainability information~\cite{eu2024825}. More broadly, the European
Green Deal, whose policy documents exhibit strong alignment with SDG~12 on sustainable
consumption and production~\cite{koundouri2024greendeal}, treats demand-side interventions
as a structural component of decarbonization rather than a regulatory afterthought.
Recommender systems occupy precisely this intersection of algorithmic design and consumption
politics: they are the primary interface through which consumers encounter products in large
online marketplaces.

In large online marketplaces, recommender systems play a central role in determining which
products become visible to users, and can therefore indirectly shape patterns of
consumption~\cite{jesse2021digitalnudging, halimeh2026nudging}. By controlling which
products a user is most likely to encounter, recommendation algorithms can potentially shift
demand toward lower-impact alternatives without restricting user choice.

Recent research has therefore begun to explore \emph{sustainability-aware recommender
systems}, where environmental signals such as carbon footprint are incorporated into
recommendation decisions~\cite{kalisvaart2025greenerrec, spillo2023sustainabilityaware,
ecoa2026, vicenti2026pcfllm}. One important sustainability signal is the \emph{Product
Carbon Footprint (PCF)}, typically measured in kilograms of CO$_2$ equivalent and estimated
using life-cycle assessment (LCA) methodologies~\cite{meinrenken2022carboncatalogue}.
Obtaining PCF values at scale is genuinely difficult: LCA studies require extensive
supply-chain data and are rarely available for the long tail of products in large
e-commerce catalogs~\cite{meinrenken2022carboncatalogue}. As a result, most
recommender-system datasets lack item-level environmental impact information.

To address this limitation, recent work has proposed estimating PCF from product metadata
using large language models (LLMs) or other predictive approaches, enabling
sustainability-aware experimentation on standard recommendation
datasets~\cite{vicenti2026pcfllm, ecoa2026}. These studies demonstrate that environmental
signals can be incorporated into recommendation pipelines, often through post-hoc
re-ranking strategies that combine predicted item relevance with estimated environmental
impact.

We study carbon-aware recommendation in a realistic e-commerce setting where PCF labels
are missing for most items. Our approach follows a modular pipeline designed to support
strong baselines and reproducible evaluation. We first generate candidate recommendations
using established collaborative filtering and neural recommendation models implemented in
the RecBole framework~\cite{recbole2021}. These models produce relevance scores that, after
per-user min-max normalisation, serve as the engagement signal in our re-ranking step. We
then apply a carbon-aware re-ranking strategy that trades off model-predicted relevance
against estimated product carbon footprint through a tunable parameter.

We evaluate this framework using the Amazon Reviews dataset as a large-scale proxy for
user and item interactions~\cite{amazon2023}, focusing on three representative product
categories: \textit{Home and Kitchen}, \textit{Sports and Outdoors}, and
\textit{Electronics}. Reviews serve as implicit signals of user engagement or purchase
behavior and are widely used in recommender-system research due to their scale and rich
product metadata~\cite{amazon2015, amazon2023, vicenti2026pcfllm, ecoa2026}. While such
observational data introduces well-known biases (including self-selection and exposure
bias), it enables controlled offline analysis of recommendation strategies at catalog scale.

Our goal is to analyze the trade-off between model-predicted relevance and the carbon
footprint of recommended products. By varying $\lambda$, we construct an engagement and
carbon Pareto frontier that characterizes how sustainability objectives interact with
recommendation performance across product categories and model families.

\paragraph{Research questions.}
We organize the study around three questions:
\begin{itemize}
    \item \textbf{RQ1:} How does carbon-aware re-ranking affect recommendation quality across different candidate generation models?
    \item \textbf{RQ2:} How much reduction in recommended-item carbon footprint can be achieved while maintaining acceptable engagement performance?
    \item \textbf{RQ3:} Do different recommendation algorithms exhibit different engagement and sustainability trade-offs?
\end{itemize}

\paragraph{Contributions.}
Prior work on sustainability-aware recommendation either assumes item-level PCF labels are
already available~\cite{spillo2023sustainabilityaware, kalisvaart2025greenerrec} or operates
on small, domain-specific datasets~\cite{vicenti2026pcfllm, ecoa2026}. We address the
realistic large-scale setting where labels are absent for nearly all catalog items. The
specific contributions are:
\begin{itemize}
  \item A \emph{retrieval-augmented PCF estimation pipeline} that transfers
    supervision from 866 life-cycle-assessed products to an unlabeled e-commerce catalog
    without requiring product-specific supply-chain data, combining few-shot LLM estimates and comparing this design against standalone
    nearest-neighbour and zero-shot LLM baselines.
  \item A \emph{multi-model Pareto analysis} across three complementary recommendation
    paradigms (pairwise collaborative filtering, neural matrix factorization, and graph
    convolution) on a large-scale public dataset, providing a unified comparison of
    engagement and carbon trade-offs in a setting where item-level PCF must be inferred.
  \item A \emph{modular, auditable re-ranking design} in which the sustainability weight
    $\lambda$ is an explicit, inspectable parameter rather than an implicit model choice,
    consistent with emerging algorithmic transparency requirements~\cite{dsa2022}.
\end{itemize}

\section{Related Work}
\label{sec:related}

\subsection{Sustainability-Aware Recommender Systems}

Recent research has begun to incorporate environmental objectives into recommender systems.
Early work demonstrates that including carbon footprint information in recommendation
ranking can reveal trade-offs between recommendation accuracy and environmental
impact~\cite{spillo2023sustainabilityaware, ecoa2026, vicenti2026pcfllm}. Similarly,
sustainability-aware recommendation frameworks have shown that post-hoc re-ranking strategies can reduce the average footprint of recommended items
while preserving recommendation quality, explicitly characterizing the trade-off between
accuracy and environmental impact~\cite{kalisvaart2025greenerrec}. These studies
establish the feasibility of incorporating environmental signals into recommender pipelines,
but typically operate on domain-specific datasets with known footprint labels. Broader
surveys of the field have further identified a systematic gap in evaluation metrics that
account for sustainability outcomes such as reduced carbon footprint, and have positioned
recommender systems as instruments for achieving the UN Sustainable Development Goals
through behavioral influence~\cite{felfernig2023sustainability}.
Kalisvaart et al.~\cite{kalisvaart2025greenerrec} further introduce a position-aware
greenness metric after mapping item-level emissions to a bounded greenness scale. In our
setting, catalog-scale Amazon PCF values are inferred rather than observed, so we report
AvgPCF@10 in kg CO$_2$e together with relative carbon reduction.

\subsection{Estimating Product Carbon Footprints}

A central challenge for sustainability-aware recommendation is the scarcity of item-level
carbon footprint data. The Carbon Catalogue provides one of the few publicly available
datasets containing product-level PCF estimates derived from life-cycle
assessments~\cite{meinrenken2022carboncatalogue}. More recent work attempts to scale
footprint estimation using machine learning approaches, including LLM-based methods that
infer PCF values from product descriptions and metadata~\cite{vicenti2026pcfllm, ecoa2026}.

Parallel advances in in-context learning have shown that large language models can perform
structured prediction tasks with minimal task-specific supervision when provided with
representative labeled examples~\cite{dong2022icl, brown2020gpt3}. Retrieval-augmented
generation extends this by grounding model outputs in retrieved evidence rather than
parametric knowledge alone, substantially improving reliability on domain-specific numeric
tasks~\cite{gao2023ragsurvey}. Asking models to reason step by step before producing a
final answer further improves numeric estimation quality~\cite{wei2022cot}.

We build on these developments by employing a retrieval-augmented estimation pipeline that
combines semantic similarity search with structured LLM prompting to infer PCF values for Amazon products from available
metadata~\cite{amazon2015, amazon2023}, enabling large-scale experimentation with
carbon-aware recommendation. Our use of retrieval-augmented prompting follows the same
general family of techniques as recent LLM-based PCF estimation work, but differs in how it
is embedded into a larger recommendation study: we compare PCF estimators on held-out
Carbon Catalogue products, select the estimator used downstream based on that benchmark,
and then evaluate how inferred PCF values affect multi-model recommendation trade-offs
rather than treating PCF estimation as the endpoint.

\subsection{Recommendation Infrastructure and Baseline Models}

Modern recommender-system research relies on standardized frameworks and strong baseline
models. RecBole provides a unified library implementing dozens of recommendation algorithms
with consistent evaluation protocols, facilitating reproducible experimentation across model
families~\cite{recbole2021}. The relevance scores produced by these models serve as direct
proxies for predicted user engagement in post-hoc re-ranking pipelines, making it
straightforward to incorporate additional objectives such as carbon footprint without
modifying the underlying recommendation model.

\subsection{Behavioral Effects of Recommendations and Digital Nudging}

Recommendation rankings influence user decisions by shaping the set of alternatives that
users consider. Research in digital nudging has systematically documented this effect,
demonstrating that recommender systems alter user behavior through mechanisms such as
framing, salience, and ranking position~\cite{jesse2021digitalnudging}. The design of
ranking criteria is therefore not a neutral technical decision; it constitutes a choice
about which products receive attention and, by extension, which consumption patterns are
reinforced. In sustainability contexts, explanations and information framing have been shown
to significantly increase the likelihood that users select environmentally friendly
products~\cite{halimeh2026nudging}. Recommendation-driven behavioral change is not always
beneficial, however: simulation studies show that when systems are trained on data already
shaped by prior recommendations, a feedback loop emerges that homogenizes user behavior
without improving utility~\cite{chaney2018confounding}. This finding underscores that the
\emph{content} of what recommendation systems amplify matters, and provides additional
motivation for introducing carbon-aware objectives as a corrective to engagement-only
optimization. Reviews spanning nudging interventions from 2008 to 2024 confirm that digital
nudging has emerged as a growing subfield, with ranking-based and salience-based
interventions among the most studied tools for promoting sustainable
choice~\cite{nudgereview2024}.

\subsection{Sustainability Evaluation Metrics}

Beyond traditional accuracy metrics, recent work proposes evaluation frameworks for
recommender systems that incorporate sustainability objectives. These include metrics based
on environmental impact, life-cycle assessment signals, and the proportion of recommended
items classified as environmentally friendly~\cite{felfernig2025sustainability}. Such work
highlights the importance of evaluating recommender systems along multiple objectives,
motivating the engagement and carbon trade-off analysis explored in this study.

\section{Background: Problem Context and Motivation}
\label{sec:background}

Integrating carbon signals into recommender systems sits at the intersection of three
pressures: the environmental footprint of e-commerce, growing policy demands for reliable
sustainability information, and the recognition that ranking systems shape consumer
choice~\cite{buldeorai2023ecommerce, koundouri2024greendeal, eu2024825,
jesse2021digitalnudging}.

These pressures matter because recommendation is a form of choice architecture. Any ranking
policy decides which products become salient, and therefore which kinds of consumption are
amplified. In sustainability settings, this makes ranking criteria a design choice rather
than a neutral implementation detail~\cite{jesse2021digitalnudging, halimeh2026nudging}.

This framing also connects to broader concerns about autonomy and accountability. Making the
trade-off explicit through $\lambda$ is more transparent than embedding a fixed
sustainability preference invisibly in model training. It also aligns with concerns about
user autonomy, platform accountability, and feedback effects in recommender
systems~\cite{bonicalzi2023autonomy, dsa2022, chaney2018confounding}.

From a technical perspective, strong recommendation backbones already exist, but item-level
PCF remains unavailable for most catalog items at scale~\cite{recbole2021,
meinrenken2022carboncatalogue}. Recent work has therefore turned to metadata-based and
LLM-based estimation to make sustainability-aware recommendation feasible in standard
catalog settings~\cite{vicenti2026pcfllm, ecoa2026}. Our framework builds on that line of
work by combining inferred PCF with transparent post-hoc re-ranking.

\section{Methodology}\label{sec:methodology}
Our approach consists of three components: (1) estimating product carbon footprints (PCF) for Amazon catalog items, (2) constructing a recommendation pipeline that combines candidate generation and carbon-aware re-ranking, and (3) evaluating the trade-off between recommendation quality and environmental impact.

\begin{figure}[H]
  \centering
  \scalebox{0.78}{\begin{tikzpicture}[node distance=0.85cm]

  \node[font=\small\bfseries, text=graytext] (hdr-l) at (0, 0)
    {PCF estimation};
  \node[font=\small\bfseries, text=graytext] (hdr-r) at (7.8, 0)
    {Recommendation pipeline};

  \node[dbnode, teal, below=0.5cm of hdr-l] (catalogue)
    {\textbf{Carbon Catalogue}\sub{866 PCF labels (LCA)}};

  \node[dbnode, amber, right=1.6cm of catalogue] (amazoncatalogue)
    {\textbf{Amazon catalogue}\sub{Metadata: title, category}};

  \node[procnode, purple, below=1.2cm of catalogue, xshift=2.0cm] (embedder)
    {\textbf{Sentence embedder}\sub{Title + description}};

  \node[procnode, purple,
        minimum height=1.3cm, text width=3.8cm, minimum width=4.0cm,
        below=0.8cm of embedder] (predictor)
    {\textbf{PCF predictor (few-shot)}\\[2pt]
     {\footnotesize\color{purpletext}
      Query: Amazon item embedding\\
      Context: 5 nearest Carbon\\
      Catalogue neighbours}};

  \node[dbnode, teal, below=0.8cm of predictor] (pcfstore)
    {\textbf{PCF store}\sub{PCF$_i$ per Amazon item}};

  \draw[solidarr]
    (catalogue.south) -- ++(0, -0.35)
    -| node[pos=0.2, below, font=\footnotesize, text=graystroke]
       {Labelled embeddings}
    (embedder.north west);

  \draw[solidarr]
    (amazoncatalogue.south) -- ++(0, -0.35)
    -| node[pos=0.2, below, font=\footnotesize, text=graystroke]
       {Query embeddings}
    (embedder.north east);

  \draw[solidarr, color=tealstroke]
    ([xshift=-0.4cm] embedder.south)
    -- node[left, font=\footnotesize, text=tealstroke] {5 nearest}
    ([xshift=-0.4cm] predictor.north);

  \draw[solidarr, color=amberstroke]
    ([xshift=0.4cm] embedder.south)
    -- node[right, font=\footnotesize, text=amberstroke] {query item}
    ([xshift=0.4cm] predictor.north);

  \draw[solidarr] (predictor) -- (pcfstore);

  \node[dbnode, amber, below=0.5cm of hdr-r] (reviews)
    {\textbf{Amazon Reviews}\sub{Implicit user--item feedback}};

  \node[procnode, gray, below=0.8cm of reviews] (dataprep)
    {\textbf{Data preparation}\sub{Train / val / test split}};

  \node[procnode, blue,
        minimum height=2.4cm, text width=4.0cm, minimum width=4.2cm,
        below=0.8cm of dataprep] (recbole)
    {\textbf{RecBole models}\\[5pt]
     {\footnotesize\ttfamily BPR \quad NeuMF \quad LightGCN}\\[4pt]
     {\footnotesize\color{bluetext} Relevance scores $\hat{y}_{ui}$}};

  \node[procnode, coral,
        minimum height=1.5cm, text width=4.0cm, minimum width=4.2cm,
        below=0.8cm of recbole] (reranker)
    {\textbf{Carbon-aware re-ranker}\\[3pt]
     {\small $s_{ui} = (1{-}\lambda)\,\hat{y}_{ui} - \lambda\,\mathrm{PCF}_i$}\\[2pt]
     {\footnotesize\color{coraltext} $\lambda \in [0,\,1]$}};

  \node[procnode, green, below=0.8cm of reranker] (topk)
    {\textbf{Top-$K$ recommendations}\sub{Final ranked list}};

  \draw[solidarr] (reviews)  -- (dataprep);
  \draw[solidarr] (dataprep) -- (recbole);
  \draw[solidarr] (recbole)  -- (reranker);
  \draw[solidarr] (reranker) -- (topk);

  \draw[crossarr]
    (pcfstore.east) -- ++(0.6, 0)
    |- node[pos=0.28, above, font=\footnotesize\bfseries, text=tealstroke]
       {PCF$_i$}
    (reranker.west);

  \coordinate (divmid) at ($(topk.south) + (0, -1.0)$);
  \draw[graystroke, dashed, line width=0.4pt, dash pattern=on 3pt off 4pt]
    ($(divmid) + (-5.0, 0)$) -- ($(divmid) + (5.0, 0)$);
  \node[font=\footnotesize\bfseries, text=graystroke, anchor=south]
    at ($(divmid) + (0, 0.08)$) {Evaluation};

  \node[procnode, blue,
        minimum width=3.2cm, text width=3.0cm,
        below=2.0cm of topk, xshift=-3.8cm] (engagement)
    {\textbf{Engagement}\sub{NDCG@10}};

  \node[procnode, teal,
        minimum width=3.2cm, text width=3.0cm,
        below=2.0cm of topk] (sustainability)
    {\textbf{Sustainability}\sub{AvgPCF@10}};

  \node[procnode, coral,
        minimum width=3.2cm, text width=3.0cm,
        below=2.0cm of topk, xshift=3.8cm] (pareto)
    {\textbf{Pareto frontier}\sub{NDCG vs AvgPCF ($\lambda$ sweep)}};

  \draw[solidarr] (topk.south) -- ++(0, -0.45) -| (engagement.north);
  \draw[solidarr] (topk.south) -- (sustainability.north);
  \draw[solidarr] (topk.south) -- ++(0, -0.45) -| (pareto.north);

\end{tikzpicture}}
  \caption{Overview of the carbon-aware recommendation framework. Product text denotes the available item metadata used for embedding and PCF estimation, primarily title and category fields in the reported Amazon run.}
  \label{fig:pipeline}
\end{figure}
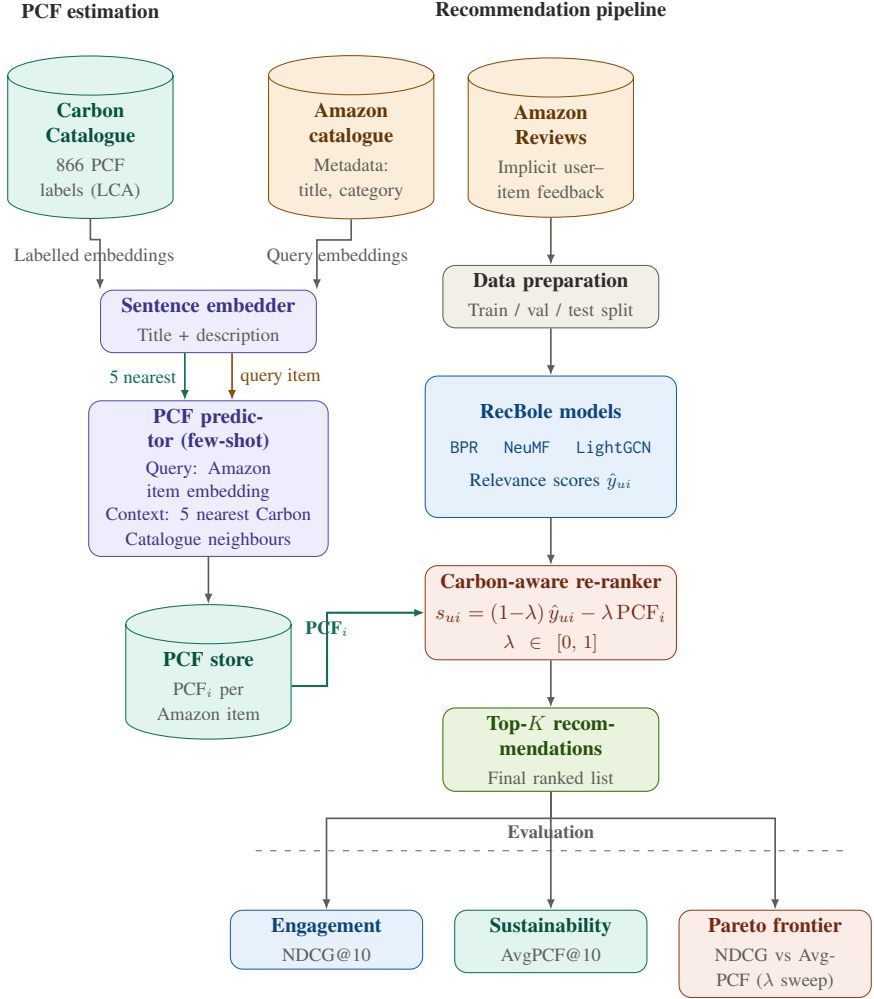

\subsection{Product Carbon Footprint Estimation}

A central challenge in sustainability-aware recommendation is that PCF labels are
unavailable for most items in large e-commerce catalogs. To address this, we estimate
product-level carbon footprints by transferring supervision from the Carbon
Catalogue~\cite{meinrenken2022carboncatalogue} to the Amazon catalog~\cite{amazon2023}. We
compare three estimation strategies of increasing sophistication on held-out Carbon
Catalogue items before applying the best-performing method downstream.

\paragraph{Nearest-neighbour average.}
As a non-parametric baseline, we embed each product title using the
\texttt{\seqsplit{all-MiniLM-L6-v2}} sentence encoder~\cite{reimers2019sbert} and retrieve the $k$ nearest
labelled neighbours from the Carbon Catalogue by cosine similarity. We treat this model as a compact sentence-embedding model rather than as a generative LLM. The PCF estimate is
the unweighted average of the neighbours' known PCF values. This baseline requires no
generation step and serves as a strong non-parametric reference point.

\paragraph{Zero-shot LLM.}
As a second baseline, we prompt an instruction-tuned LLM
(\texttt{\seqsplit{Qwen/Qwen2.5-3B-Instruct}}) with the product title alone,
without any retrieved examples, and ask it to estimate PCF in kg\,CO$_2$e. The prompt
specifies a typical PCF range (1--10{,}000 kg CO$_2$e) and a strict output format (one
number, no scientific notation); parsed values are clamped to a plausible range before
evaluation. This tests whether parametric world knowledge encoded during pre-training is
sufficient for numeric carbon estimation, without any grounding in labeled product examples.

\paragraph{Few-shot retrieval-augmented LLM.}
Our primary estimator combines semantic retrieval with few-shot
prompting~\cite{brown2020gpt3}. For each query product, we embed its title using the
Sentence-Transformers model \texttt{\seqsplit{all-MiniLM-L6-v2}}~\cite{reimers2019sbert} and retrieve its five nearest Carbon
Catalogue neighbours by cosine similarity. These neighbours are assembled into a structured
prompt in which they serve as labelled in-context examples; the full prompt template is provided in
Figure~\ref{fig:llm-prompt-appendix} in the appendix. The model is instructed to reason step by step before
producing a final PCF estimate~\cite{wei2022cot}, following the retrieval-augmented
estimation strategy of Vicenti et al.~\cite{vicenti2026pcfllm}.

\paragraph{Evaluation and method selection.}
We evaluate all three methods on a held-out slice of 866 Carbon Catalogue items, hiding the
true PCF and scoring predictions by RMSE, MAE, and Spearman rank correlation. We then choose a single downstream PCF source based on this estimator benchmark. In the reported run, the few-shot retrieval-augmented LLM has the lowest consumer-scale RMSE, MAE, and median absolute error, so it is selected as the main PCF signal for Amazon re-ranking. For each Amazon catalog item, we embed its title, retrieve the
five nearest Carbon Catalogue neighbours, prompt the LLM with those neighbours as labelled calibration examples, and retain the resulting product-level estimate
$\mathrm{PCF}_i$. Items with insufficient metadata for embedding are excluded from
downstream analysis. In the reported run, the full Amazon catalog is scored with
\texttt{Qwen/Qwen2.5-3B-Instruct}; only the small number of products for which the few-shot output is invalid fall back to the nearest-neighbour estimate, as described in Appendix~\ref{appendix:A}.

\subsection{Recommendation Pipeline}

Our recommendation pipeline is modular and consists of three stages.

\paragraph{Data preparation.}
We process the Amazon Reviews dataset to construct user and item interaction histories. Each
review is treated as implicit feedback: it does not prove that a product was purchased or liked, but it provides an observable signal that the user interacted with the product and is therefore used as our offline proxy for engagement. Interactions are split into
training, validation, and test sets by timestamp, ensuring each user's held-out
interactions occur strictly after their training history. After filtering and RecBole formatting, the category-specific datasets contain tens of
thousands of users, 24{,}000 items per category, and between 535{,}000 and
1.04 million interactions (Table~\ref{tab:dataset-stats}). PCF estimates are
assigned to items from a catalog of 72{,}000 scored Amazon products; items without a PCF
estimate or user interaction are excluded from re-ranking.

\paragraph{Candidate generation.}
For each user in the test set, we generate a candidate set of top-$K$ recommendations
using three established algorithms implemented in the RecBole framework~\cite{recbole2021}:
BPR, NeuMF, and LightGCN. Each model is trained on the training split and produces a
relevance score $\hat{y}_{ui}$ for user $u$ and item $i$, which serves as our proxy for
predicted engagement. Items are ranked by $\hat{y}_{ui}$, and the top-$K$ items are
retained as candidate recommendations.

\paragraph{Carbon-aware re-ranking.}
We re-rank each candidate list by combining a normalised engagement score with a normalised
carbon footprint. Let $\tilde{y}_{ui} \in [0,1]$ denote the relevance score
$\hat{y}_{ui}$ after per-user min-max normalisation, and let $\widetilde{\mathrm{PCF}}_i
\in [0,1]$ denote the estimated product carbon footprint after global min-max normalisation
across all catalog items. For each candidate pair $(u,i)$, the final ranking score is

\begin{equation}
s_{ui} = (1 - \lambda)\, \tilde{y}_{ui} - \lambda \, \widetilde{\mathrm{PCF}}_i,
\end{equation}

where $\lambda \in [0,1]$ controls the trade-off between engagement and sustainability.
Normalising both signals before combining them ensures that $\lambda$ has a consistent and
comparable effect across users and items regardless of the raw score scales produced by
different recommendation models. This choice also makes the endpoints interpretable: $\lambda=0$ recovers the original model ranking, while $\lambda=1$ ranks candidates only by the normalized carbon signal. Because min--max normalization can be sensitive to outliers, we treat alternative normalizations such as z-scoring or rank-based scaling as useful robustness checks for future work. 

This formulation corresponds to a linear scalarization of a two-objective optimisation
problem, in which relevance is maximized and carbon footprint is minimized. The linear form
provides a transparent and interpretable trade-off, allowing $\lambda$ to directly control
the marginal substitution between engagement and environmental impact. While more complex
non-linear combinations are possible, we adopt this formulation to ensure that the effect of
$\lambda$ is monotonic and easy to audit.

When $\lambda = 0$, ranking depends only on model-predicted relevance; when $\lambda = 1$,
ranking prioritizes lower-carbon items.

\subsection{Evaluation and Trade-off Analysis}

We evaluate the carbon-aware recommendation lists using both engagement and sustainability
metrics.

\paragraph{Engagement metrics.}
We use NDCG@10 as the primary engagement metric. NDCG is well-suited to our setting
because it rewards placing the held-out relevant item higher in the top-10 list, which is
exactly the behavior that carbon-aware re-ranking changes. While NDCG@10 captures ranking quality, it is computed under a leave-one-out
protocol with a single held-out item per user and therefore does not fully
represent user utility in settings with multiple relevant items. The metric is
appropriate for controlled offline comparison, but future work should validate
these findings using richer relevance signals and online evaluation. For context, a random ranking
over the item set yields an expected NDCG@10 on the order of $10^{-4}$, indicating that all
evaluated models perform substantially above chance.\footnote{Under a uniform random ranking over $N$ items with a single relevant item,
the probability that the item appears at rank $i$ is $1/N$. The expected NDCG@10 is
$\mathbb{E}[\mathrm{NDCG@10}] = \frac{1}{N} \sum_{i=1}^{10} \frac{1}{\log_2(i+1)}$,
which evaluates to approximately $1.9 \times 10^{-4}$ for $N \approx 24{,}000$,
consistent with the scale of our item sets.}
We do not emphasize Recall@10 because, under leave-one-out evaluation with a single
held-out item per user, it reduces to a hit rate at 10 and does not capture ranking quality.
\paragraph{Sustainability metric.}
We measure environmental impact using AvgPCF@10, defined as the average predicted product
carbon footprint of items appearing in each user's top-10 recommendation list.
This differs from prior work such as Kalisvaart et al.~\cite{kalisvaart2025greenerrec},
which evaluates list greenness after transforming emissions into a bounded greenness
score. We instead keep the metric in kg CO$_2$e. This is easier to interpret and matches
the quantity penalized by the re-ranker. We also report relative carbon reduction to
compare trade-offs across models and categories.

\paragraph{Carbon reduction.}
To provide an interpretable measure of environmental improvement, we additionally report
the percentage reduction in recommended-item carbon footprint relative to the
engagement-only baseline ($\lambda=0$):

\begin{equation}
\mathrm{Reduction} =
\frac{\mathrm{AvgPCF}_{\lambda=0} - \mathrm{AvgPCF}_{\lambda}}
{\mathrm{AvgPCF}_{\lambda=0}}.
\end{equation}

\paragraph{Pareto frontier.}
To characterize the engagement and sustainability trade-off, we vary $\lambda$ over a fixed
grid and compute NDCG@10 and AvgPCF@10 for each setting. We then plot NDCG@10
against AvgPCF@10 to obtain a Pareto frontier, identifying operating points for which no
other configuration simultaneously achieves higher engagement and lower carbon footprint.

\section{Experiments}\label{sec:experiments}
We evaluate the proposed framework on the Amazon Reviews dataset as a large-scale proxy for
user and item interactions. The goal is to quantify how much environmental impact can be
reduced through carbon-aware re-ranking while preserving recommendation quality.

\subsection{Experimental Setup}

We follow the pipeline described in Section~4. Candidate generation models are trained on
the timestamp-ordered training split and scored on held-out test users. Evaluation is
performed on top-10 recommendation lists over category-specific item sets
(24{,}000 items per category; Table~\ref{tab:dataset-stats}).

We generate top-$K$ candidate sets (with $K=100$) for each user and apply carbon-aware
re-ranking within this candidate pool. The candidate-pool size is important because re-ranking can only substitute among items already retrieved by the base recommender: a larger pool such as $K=500$ or $K=1000$ could expose more low-carbon alternatives, but would also change the computational cost and potentially the relevance distribution available to the re-ranker. We use $K=100$ as a pragmatic setting that gives the re-ranker substantially more than the final top-10 list while keeping full-sort scoring and per-user re-ranking tractable across all model--category combinations. The resulting top-10 lists are evaluated using the
held-out user--item interactions from the test split.

$\lambda$ is swept over a 25-point grid spanning $[0,1]$. The sweep uses smaller increments near the extremes
where the trade-off changes most rapidly: increments of 0.025 for $\lambda \leq 0.1$,
increments of 0.05--0.1 between $\lambda=0.1$ and $\lambda=0.8$, and increments of 0.01
from $\lambda=0.90$ to $\lambda=1.00$ to resolve the engagement cliff.

\subsection{Candidate Generation Baselines}

To test whether the effects of carbon-aware re-ranking are robust across recommendation
paradigms, we use three standard RecBole models as candidate generators:

\begin{itemize}
    \item \textbf{BPR} (Bayesian Personalized Ranking), a pairwise collaborative filtering model for implicit feedback;
    \item \textbf{NeuMF} (Neural Matrix Factorization), a neural hybrid of matrix factorization and multi-layer perceptrons;
    \item \textbf{LightGCN} (Light Graph Convolutional Network), a graph-based collaborative filtering model defined over the user and item interaction graph.
\end{itemize}

These baselines span collaborative, neural, and graph-based recommendation families.

\subsection{Evaluation Protocol}

For each candidate generation baseline, we apply carbon-aware re-ranking over a sweep of
$\lambda$ values and evaluate the resulting top-10 recommendation lists using NDCG@10,
AvgPCF@10, and carbon reduction relative to the $\lambda=0$ baseline. This
protocol allows us to measure both the recommendation cost and the environmental benefit of
increasing the weight placed on carbon footprint.

\subsection{Connection to Research Questions}

The experimental protocol directly operationalizes the research questions introduced in Section~\ref{sec:introduction}: the $\lambda$ sweep measures quality degradation and carbon reduction (RQ1--RQ2), while repeating the same procedure for BPR, NeuMF, and LightGCN enables cross-model comparison (RQ3).

\subsection{Trade-off Visualization}

To analyze the effect of carbon-aware re-ranking, we plot NDCG@10 against AvgPCF@10 for
each value of $\lambda$ and for each candidate generation model. These curves define Pareto
frontiers that summarize the achievable trade-off between engagement and environmental
impact. We additionally report carbon reduction percentages to quantify the practical
benefit of moving away from the engagement-only baseline.

\section{Results}\label{sec:results}

\subsection{PCF Estimation Quality}
\label{sec:pcf-results}

Before evaluating the downstream trade-off, we assess the quality of the PCF estimation
module. The environmental signal is only as reliable as the estimator that produces it.

\paragraph{Evaluation setup.}
We evaluate all three estimation methods on a held-out slice of Carbon Catalogue products
(seed fixed for reproducibility). For each Carbon Catalogue item, we hide its PCF, exclude
it from retrieval, and predict from the five nearest remaining labelled neighbours. Zero-shot
and few-shot prompts include a typical PCF scale and strict output format; outputs are
clamped before evaluation. We report RMSE, MAE, median absolute error, and Spearman rank
correlation. The Carbon Catalogue contains 866 products across multiple sectors, including both ordinary consumer products and a small heavy-industrial tail. To keep the benchmark aligned with the consumer-product setting studied
downstream, we treat the \emph{consumer-scale} subset with true PCF $\leq
10{,}000$~kg CO$_2$e as the primary scope. This removes 92 heavy-tail products, leaving 774 consumer-scale items. We then report metrics on the
\emph{intersection} of that subset with items where \emph{all} base estimators returned a
valid prediction ($n=771$), so every method is evaluated on the same rows. The remaining three consumer-scale products are excluded only because at least one LLM baseline did not produce a parseable numeric answer under our strict output parser. Constrained decoding or function-style structured generation would likely remove this failure mode. The downstream
pipeline uses few-shot LLM for predicting PCF.

\begin{table}[H]
\centering
\small
\begin{tabular}{lrrrrr}
\hline
Method & RMSE & MAE & Median AE & Spearman $\rho$ \\
\hline
Neighbour average & 3{,}002.0 & 959.4 & 123.3 & \textbf{0.728} \\
Zero-shot LLM & 1{,}712.8 & 790.7 & 93.2 & 0.064 \\
Few-shot LLM & \textbf{1{,}708.6} & \textbf{695.1} & \textbf{58.6} & 0.518 \\
\hline
\end{tabular}
\caption{PCF estimation on the consumer-scale hold-out intersection
($\mathrm{PCF}_{\mathrm{true}} \leq 10{,}000$~kg CO$_2$e, all methods
valid; $n=771$). Lower RMSE, MAE, and median absolute error are better;
higher Spearman $\rho$ is better.}
\label{tab:pcf}
\end{table}

On the consumer-scale benchmark, the few-shot LLM is the strongest estimator on absolute
error, achieving the lowest RMSE, MAE, and median absolute error. Because the downstream
re-ranker depends on reasonably calibrated absolute PCF values rather than ranking alone,
we use few-shot LLM as the PCF signal. The neighbour-average baseline remains best on
rank preservation ($\rho=0.728$), indicating that retrieval alone captures useful local
ordering even when its absolute values are less accurate.

The full 866-item holdout remains a robustness check. There, the neighbour-average
baseline performs best because a small number of heavy industrial products dominate the
error metrics: 92 held-out items have true PCF above 10{,}000~kg CO$_2$e, beyond the
consumer-oriented clamp used in the LLM prompts. The appendix diagnostics
(Figures~\ref{fig:pcf-subset-accuracy}--\ref{fig:pcf-sector-tail}) show that this heavy
tail is concentrated in a few sectors and is not representative of the downstream Amazon
recommendation regime.

\subsection{Main Trade-off Pattern}
\label{sec:lambda-results}

\begin{figure}[H]
  \centering
  \begin{subfigure}[t]{0.32\textwidth}
    \includegraphics[width=\textwidth]{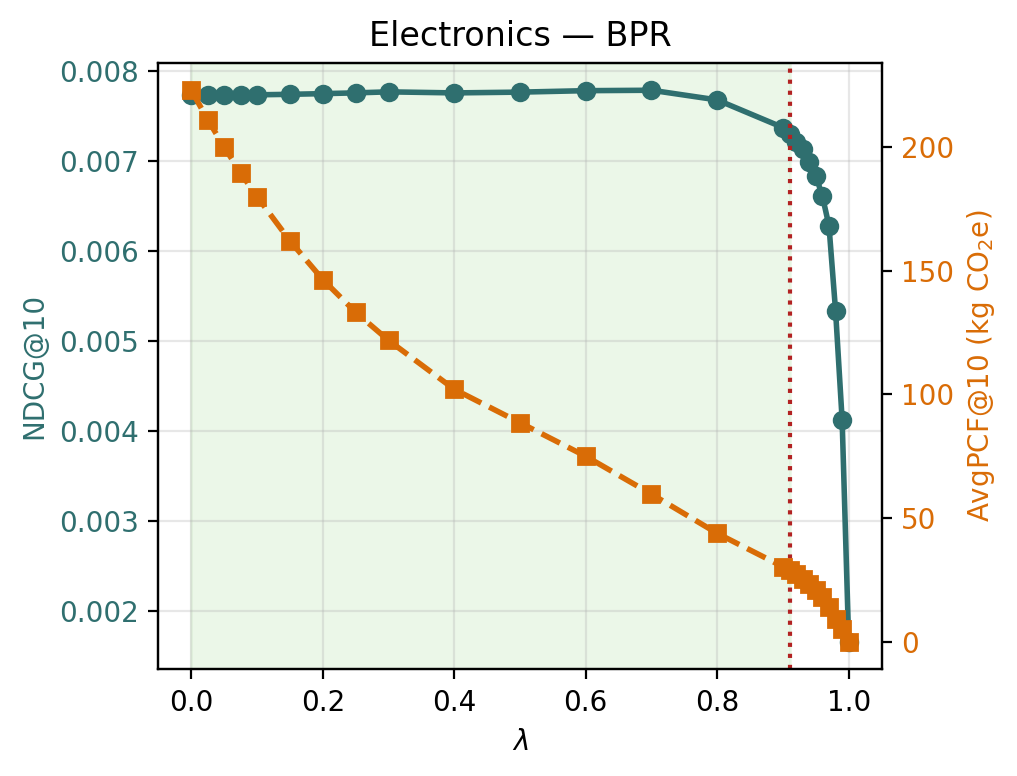}
    \caption{BPR.}
  \end{subfigure}
  \hfill
  \begin{subfigure}[t]{0.32\textwidth}
    \includegraphics[width=\textwidth]{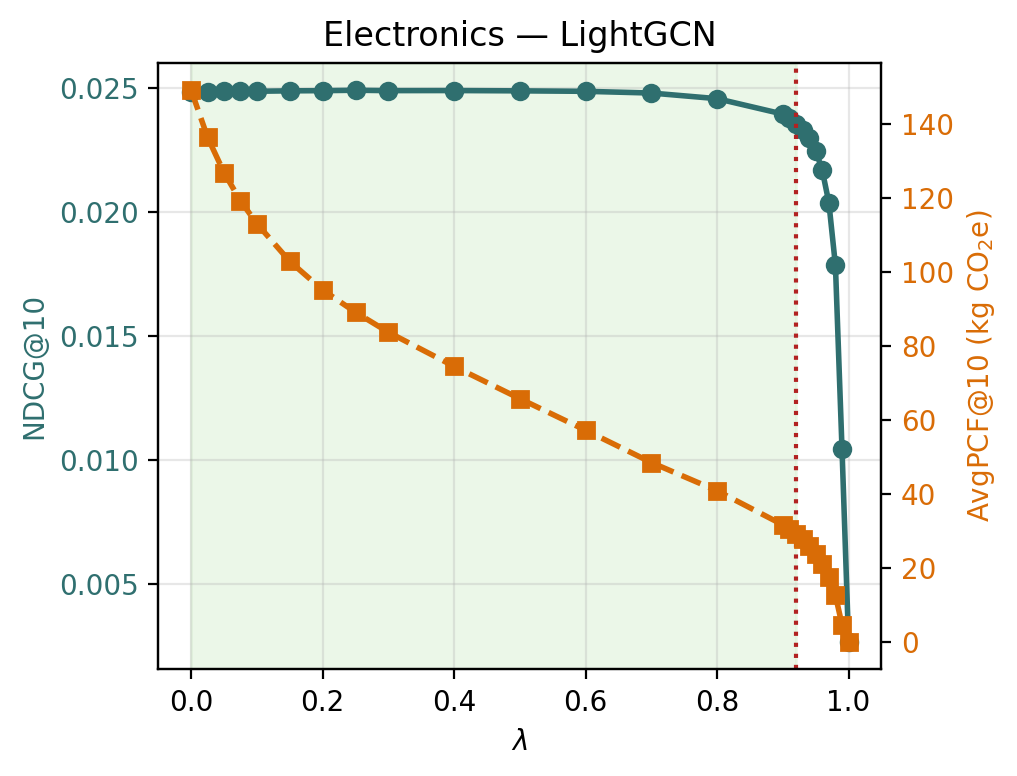}
    \caption{LightGCN.}
  \end{subfigure}
  \hfill
  \begin{subfigure}[t]{0.32\textwidth}
    \includegraphics[width=\textwidth]{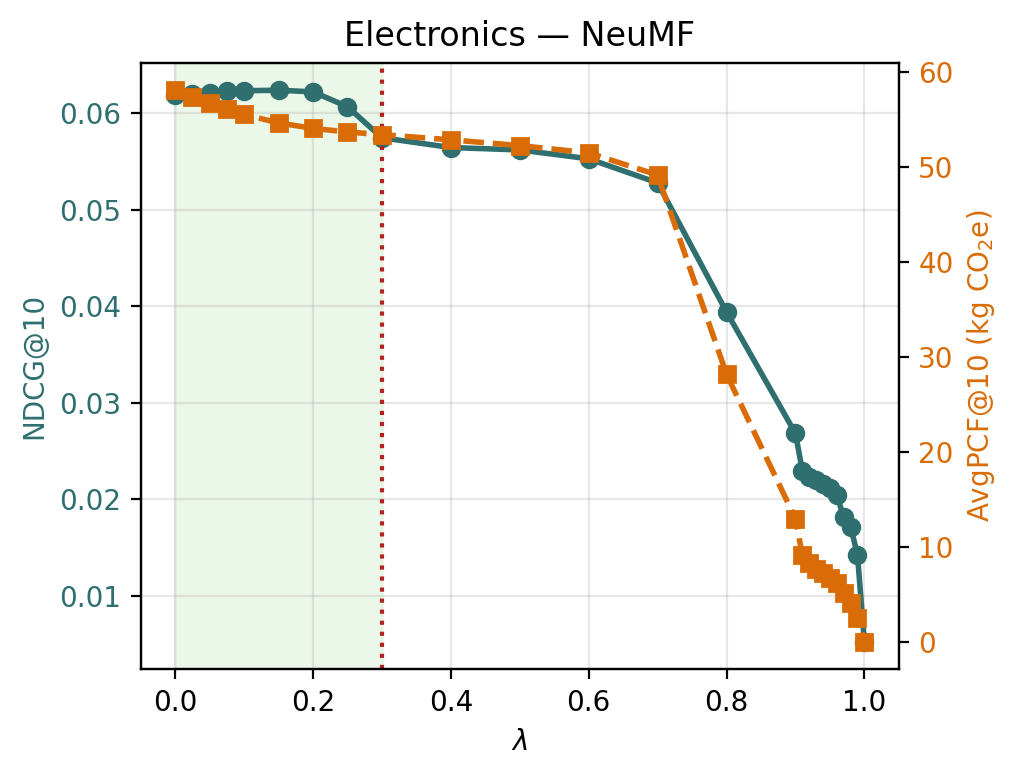}
    \caption{NeuMF.}
  \end{subfigure}
  \caption{$\lambda$ sensitivity for the Electronics category. Each panel plots
    NDCG@10 (solid, left axis) and AvgPCF@10 (dashed, right axis) as
    $\lambda$ increases from the engagement-only baseline
    ($\lambda=0$) toward carbon-only re-ranking ($\lambda=1$). The shaded green region marks operating points whose NDCG@10 remains within 5\% of the $\lambda=0$ baseline.}
  \label{fig:lambda-electronics}
\end{figure}

We first examine how NDCG@10 and AvgPCF@10 evolve as $\lambda$ increases. Figure~\ref{fig:lambda-electronics}
shows Electronics as a representative example; the corresponding Home \& Kitchen and Sports
\& Outdoors curves are provided in the appendix (Figures~\ref{fig:lambda-home}
and~\ref{fig:lambda-sports}).

The dominant pattern is a plateau followed by a sharp decline. For BPR and LightGCN, NDCG@10
remains nearly flat through most of the $\lambda$ grid while AvgPCF@10 falls steadily.
Under a 5\% NDCG budget, these two models typically achieve roughly 70--86\% carbon
reduction across the three categories.

NeuMF exhibits less flexibility under re-ranking. Its higher baseline engagement comes with a more peaked score
distribution, so re-ranking disrupts relevance earlier, most clearly in Electronics where
only 7.6\% carbon reduction fits within the same 5\% budget. Home \& Kitchen is the
exception: there, moderate $\lambda$ values improve both NDCG@10 and carbon footprint.
We treat that effect as category-specific rather than general. Overall, substantial carbon
savings are often available before recommendation quality collapses, but the size of this
region depends strongly on the backbone model.

\subsection{Cross-Model Comparison}
\label{sec:cross-model}

\begin{figure}[H]
  \centering
  \begin{subfigure}[t]{0.32\textwidth}
    \includegraphics[width=\textwidth]{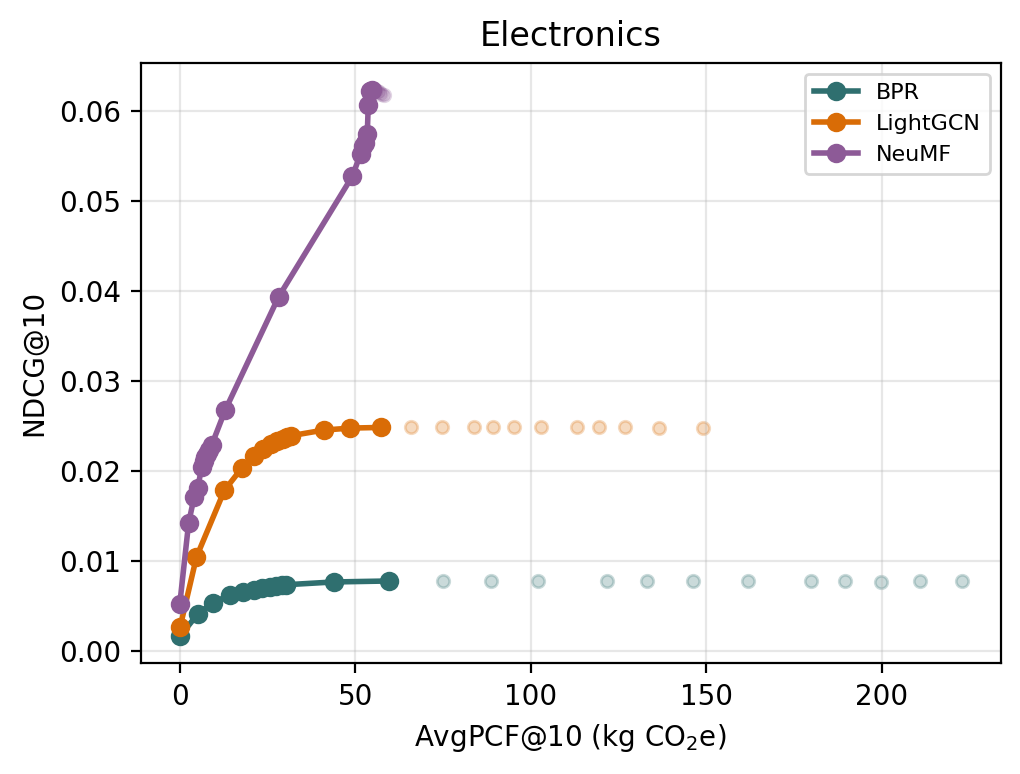}
    \caption{Electronics.}
  \end{subfigure}
  \hfill
  \begin{subfigure}[t]{0.32\textwidth}
    \includegraphics[width=\textwidth]{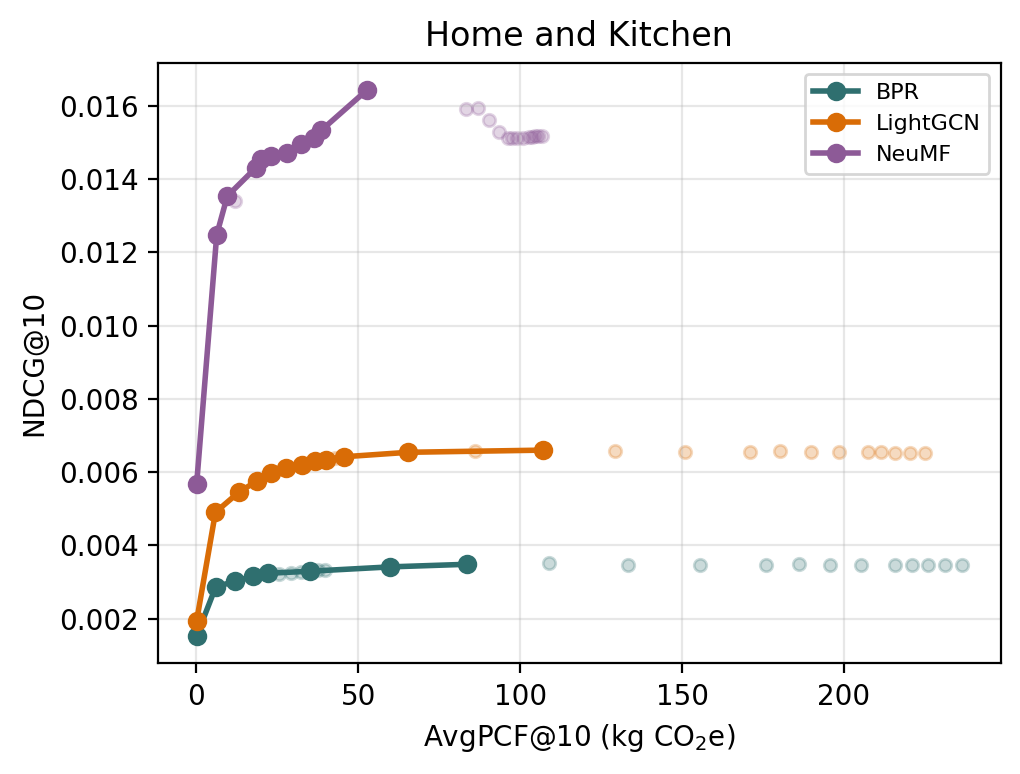}
    \caption{Home and Kitchen.}
  \end{subfigure}
  \hfill
  \begin{subfigure}[t]{0.32\textwidth}
    \includegraphics[width=\textwidth]{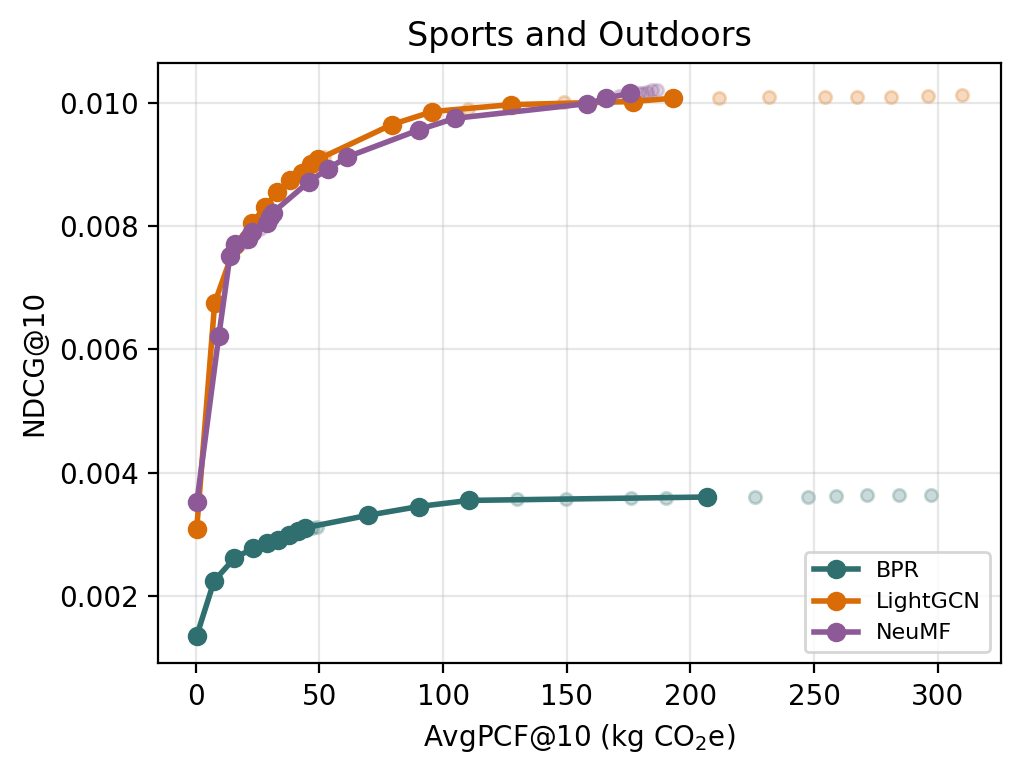}
    \caption{Sports and Outdoors.}
  \end{subfigure}
  \caption{Multi-model Pareto frontier comparison. Each panel overlays the
    Pareto-optimal trade-offs of BPR, LightGCN, and NeuMF for one
    category; curves closer to the upper-left corner offer a better
    engagement and carbon trade-off. Faint points show evaluated $\lambda$ settings that are dominated, while the darker connected points show the Pareto-optimal subset.}
  \label{fig:multimodel-pareto}
\end{figure}

We use Pareto frontiers to compare the non-dominated engagement and carbon operating points
across models. Figure~\ref{fig:multimodel-pareto} overlays the Pareto-optimal frontiers of
all three recommenders within each category. Additional per-category Pareto plots are provided in the appendix
(Figures~\ref{fig:pareto-electronics}--\ref{fig:pareto-sports}).

NeuMF achieves the highest absolute engagement in all three categories, so it remains the
strongest choice when preserving NDCG is the primary objective. BPR and LightGCN, however,
usually expose more carbon headroom before quality degrades. Within a 5\% NDCG budget, BPR
leads in Electronics and Home \& Kitchen, while LightGCN leads in Sports \& Outdoors.
Model choice therefore affects both baseline quality and the shape of the trade-off.

\subsection{Category-Level Variation}
\label{sec:variation-results}

\begin{figure}[H]
  \centering
  \includegraphics[width=0.65\textwidth]{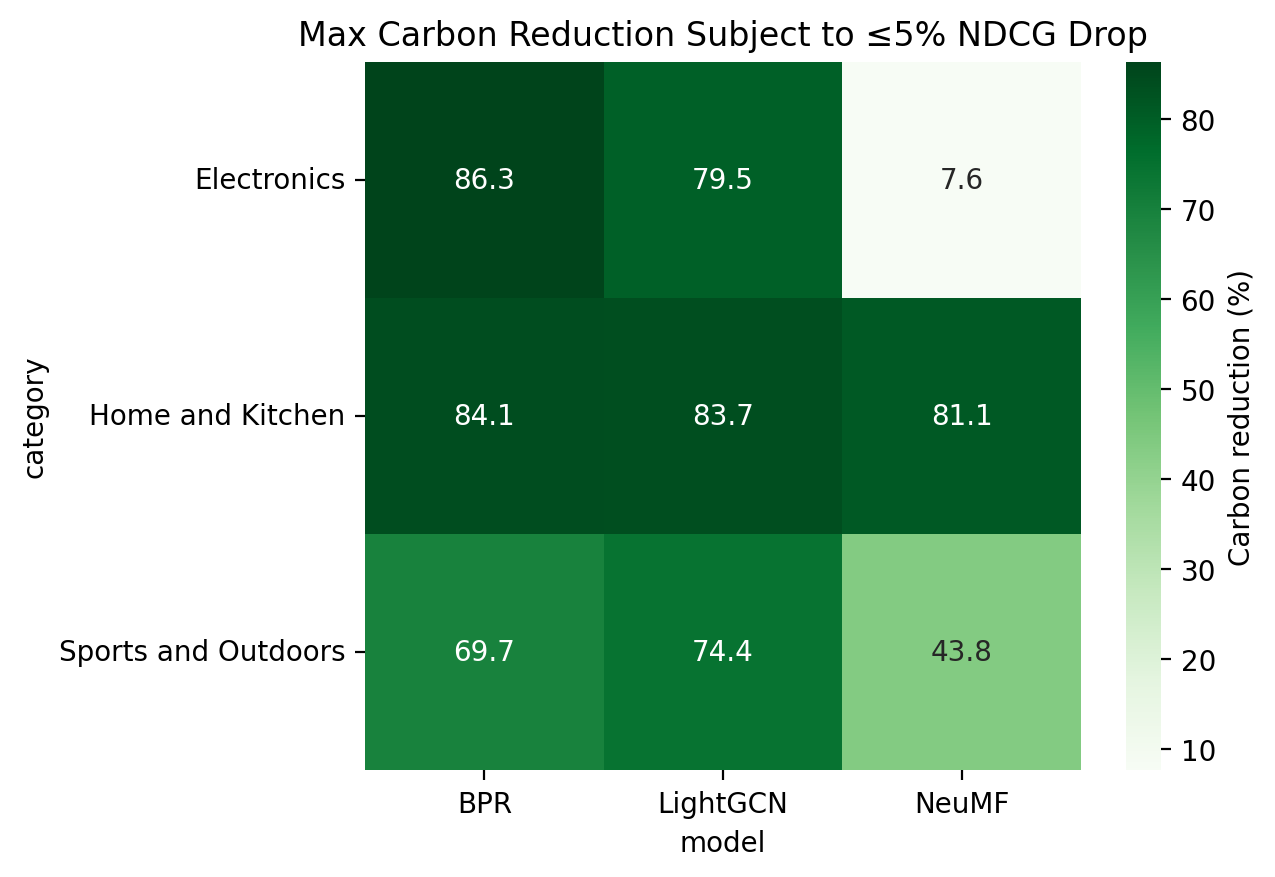}
  \caption{Heatmap of maximum carbon reduction (\%) achievable while
    limiting NDCG@10 degradation to at most 5\% relative to the
    $\lambda=0$ baseline, across all model and category combinations.}
  \label{fig:carbon-heatmap}
\end{figure}

The achievable trade-off also depends on the product category. Figure~\ref{fig:carbon-heatmap}
summarises the maximum carbon reduction achievable while limiting NDCG@10 degradation to at
most 5\% relative to the $\lambda=0$ baseline; the corresponding cross-category frontier
overlays are moved to the appendix (Figure~\ref{fig:cross-category-pareto}).

Category-level variation is substantial. Home \& Kitchen is the most carbon-flexible
category overall: all three models exceed 80\% reduction within the 5\% NDCG budget.
Electronics shows the strongest model dependence, ranging from 86.3\% reduction for BPR to
7.6\% for NeuMF. Sports \& Outdoors lies between these extremes.

These differences suggest that the trade-off depends on both category structure and model
behavior. Categories such as Home \& Kitchen appear to offer more low-carbon substitutes
with similar utility, while Electronics is more restrictive. A single global $\lambda$ is therefore unlikely to transfer cleanly across both models and categories.

One useful interpretation is in terms of \emph{substitutability} and \emph{score
sharpness}. Categories with many close substitutes allow the re-ranker to replace
higher-carbon items at low relevance cost. NeuMF, by contrast, tends to produce sharper
relevance scores. That makes its rankings more sensitive to perturbation and helps explain
why its carbon frontier often collapses earlier than BPR and LightGCN, especially in
Electronics.

\section{Limitations and Future Work}\label{sec:limitations}
Our framework enables systematic analysis of engagement and sustainability trade-offs in
recommender systems, but several limitations deserve acknowledgment.

\paragraph{Observational data biases.}
Our experiments rely on the Amazon Reviews dataset as a proxy for user and item
interactions. While widely used in recommender-system research, review datasets introduce
several well-known biases. Only a small fraction of purchases result in reviews, leading to
\emph{selection bias} where reviewers may not represent the broader user population. The
absence of a review does not reveal whether the user disliked the item, liked it but never
chose to review it, or simply had no exposure to it, creating the standard implicit-feedback
problem of ambiguous unobserved interactions. Review data may also be affected by \emph{temporal
confounds}, including changes in product popularity, seasonality, and shifting consumer
trends. These limitations are inherent to many offline recommender-system benchmarks and
may affect the interpretation of engagement metrics.

\paragraph{Carbon footprint estimation uncertainty.}
Our approach relies on predicted
product carbon footprints derived from metadata and external PCF datasets.
Although the few-shot LLM estimator performs well on the consumer-scale
benchmark, the absence of ground-truth PCF labels for the Amazon catalog means
that downstream results depend on predicted rather than observed environmental
impact. This introduces uncertainty into the reported trade-offs, and errors in
PCF estimation may propagate into the re-ranking stage. A further robustness issue is that the re-ranker uses global min--max normalized PCF values; alternative scaling choices such as z-scores, quantile normalization, or rank-based carbon penalties could change the shape of the frontier when the PCF distribution is heavy-tailed.

\paragraph{Offline evaluation limitations.}
The experiments are conducted using offline recommendation evaluation, which cannot fully
capture how users would respond to sustainability-aware recommendations in a real-world
setting. In practice, user behavior may change when exposed to environmentally informative
signals, explanations, or interface changes. Online A/B testing or controlled behavioral experiments are a natural next step to evaluate how users respond to sustainability-aware recommendations in real-world settings, and to assess whether the observed offline trade-offs translate into measurable changes in purchasing behavior.

\paragraph{Computational sustainability trade-offs.}
An additional consideration is the environmental footprint of the machine learning systems
themselves. The computational cost of models used for PCF estimation and candidate
generation is typically small compared to the life-cycle emissions of physical products,
but large-scale recommendation systems may still incur non-negligible energy consumption.
Understanding the relationship between the environmental benefits of carbon-aware
recommendations and the computational footprint of the underlying models remains an
important direction for future work.

\paragraph{Future work.}
Several directions could extend the present study. First, future work could evaluate
carbon-aware recommendation using richer behavioral datasets that contain full browsing and
purchase histories rather than relying on review-based proxies. Second, improving PCF
estimation methods, for example through structured supply-chain data, more powerful
domain-specific models, or larger LLMs, could reduce uncertainty in sustainability signals. Third, integrating
carbon-aware objectives directly into recommendation model training instead of applying
post-hoc re-ranking may yield more efficient multi-objective recommendation strategies.
Finally, online experiments and user studies could investigate how sustainability-aware
recommendations influence real purchasing decisions and whether transparency or explanation
mechanisms increase user acceptance of environmentally informed rankings. Additional offline robustness checks should also vary the candidate-pool size used for re-ranking, for example comparing $K=100$ with $K=500$ or $K=1000$, to test how much of the observed carbon headroom comes from the retrieval breadth of the base recommender.

\section{Discussion}
\label{sec:discussion}

Our main result is that carbon-aware re-ranking exhibits a broad low-cost region: for many
$\lambda$ values, AvgPCF@10 drops substantially before NDCG@10 declines sharply. This makes
post-hoc re-ranking attractive because it can reduce carbon exposure without retraining the
base model.

The attainable trade-off is not uniform. NeuMF has the strongest baseline engagement but
usually less carbon flexibility, whereas BPR and LightGCN expose more headroom. The same is
true across categories: Home \& Kitchen is consistently more flexible than Electronics.
Sustainability-aware recommendation is therefore not only a re-ranking problem, but also a
model-selection and category-structure problem.

These findings are broadly consistent with prior work showing that re-ranking can improve
sustainability at limited accuracy cost~\cite{kalisvaart2025greenerrec}. The main
difference is that our setting requires inferred rather than observed PCF values. The
trade-off remains visible even under that additional uncertainty, which suggests that the
approach is robust enough to study in large e-commerce catalogs where direct PCF labels are
missing.

Finally, $\lambda$ functions as an explicit policy lever. It exposes the engagement--sustainability trade-off directly, making the system easier to inspect, tune, and audit than approaches where the sustainability objective is embedded implicitly in model training.

\section{Conclusion}
\label{sec:conclusion}

We studied carbon-aware product recommendation in the realistic setting where Product
Carbon Footprint (PCF) labels are unavailable for most catalog items and must be inferred.
We combined retrieval-augmented PCF estimation with a post-hoc re-ranking strategy
controlled by a single parameter $\lambda$.

Across three recommendation models and three product categories, substantial reductions in
recommended-item carbon footprint are often available before recommendation quality drops
sharply. The size of that region, however, depends strongly on both the model and the
category.

These results suggest that sustainability-aware recommendation is feasible but context-dependent. Model choice and category structure both shape the achievable trade-off.

Because $\lambda$ is explicit and tunable, the framework offers a transparent way to add
sustainability objectives to existing recommendation pipelines. This makes carbon-aware
re-ranking a practical mechanism for studying and deploying sustainability trade-offs in
large e-commerce recommendation. That transparency is also consistent with emerging
accountability expectations for recommender systems~\cite{dsa2022}.

More broadly, the results support a simple conclusion: recommendation systems can influence
the carbon profile of what users are exposed to, and even a lightweight re-ranking
intervention can make that influence measurable and controllable.

\bibliographystyle{plain}
\bibliography{references}

\newpage
\begin{appendices}

\section{Hyperparameter and Experimental Configuration}\label{appendix:A}

\subsection*{RecBole Model Configurations}

All three recommendation models are trained with a unified set of hyperparameters
using the RecBole framework~\cite{recbole2021}. The configuration is identical across
models except where noted.

\begin{table}[H]
\centering
\small
\begin{tabular}{lll}
\hline
Hyperparameter & Value & Notes \\
\hline
Embedding size          & 64       & All models \\
Epochs                  & 50       & All models \\
Training batch size     & 8{,}192  & Colab run override used for reported results \\
Learning rate           & 0.001    & All models \\
Negative sampling       & uniform (1:1) & All models \\
Validation frequency    & every 10 epochs & Colab run override (`eval\_step=10`) \\
Early stopping patience & 10 validation checks & Measured on NDCG@10 \\
Evaluation mode         & full sort & Ranked against all items \\
Split strategy          & TS: [0.8, 0.1, 0.1] & Timestamp-ordered train/valid/test split \\
Evaluation batch size   & 16{,}384 & Colab run override used for reported results \\
MLP hidden layers (NeuMF) & [128, 64, 32] & NeuMF only \\
Dropout (NeuMF)         & 0.1      & NeuMF only \\
Graph convolution layers (LightGCN) & 3 & LightGCN only \\
\hline
\end{tabular}
\caption{RecBole training and evaluation hyperparameters.}
\label{tab:recbole-config}
\end{table}

\subsection*{Re-ranking Configuration}

\begin{table}[H]
\centering
\small
\begin{tabular}{ll}
\hline
Parameter & Value \\
\hline
Candidate pool size from RecBole & 100 \\
Final re-ranked list size & 10 \\
$\lambda$ grid & 0.0, 0.025, 0.05, 0.075, 0.1, 0.15, 0.2, 0.25, \\
               & 0.3, 0.4, 0.5, 0.6, 0.7, 0.8, 0.9, 0.91, 0.92, \\
               & 0.93, 0.94, 0.95, 0.96, 0.97, 0.98, 0.99, 1.0 \\
Engagement normalisation & Per-user min-max to $[0,1]$ \\
Carbon normalisation     & Global min-max to $[0,1]$ \\
User scoring batch size  & 1{,}024 \\
Missing PCF fill         & Median of available item PCFs \\
\hline
\end{tabular}
\caption{Carbon-aware re-ranking configuration.}
\label{tab:reranking-config}
\end{table}

\subsection*{PCF Estimation Configuration}

\begin{table}[H]
\centering
\small
\begin{tabular}{ll}
\hline
Parameter & Value \\
\hline
Sentence encoder        & \texttt{sentence-transformers/all-MiniLM-L6-v2} \\
Embedding dimension     & 384 \\
Similarity metric       & Cosine similarity \\
Neighbours retrieved    & 5 \\
LLM used in the reported run & \texttt{Qwen/Qwen2.5-3B-Instruct} \\
Evaluation rows         & 866 total  \\
Downstream selected PCF & Few-shot \\
PCF clamp range         & [0.01, 10{,}000] kg CO$_2$e \\
Zero-shot format        & One number, no scientific notation \\
Few-shot reasoning      & Chain-of-thought before final estimate \\
\hline
\end{tabular}
\caption{PCF estimation pipeline configuration.}
\label{tab:pcf-config}
\end{table}

\subsection*{Dataset Statistics}

\begin{table}[H]
\centering
\small
\begin{tabular}{lrrrr}
\hline
Category & Users & Items & Interactions & Avg.\ interactions/user \\
\hline
Electronics & 110{,}550 & 24{,}000 & 1{,}036{,}192 & 9.4 \\
Home \& Kitchen & 91{,}421 & 24{,}000 & 874{,}971 & 9.6 \\
Sports \& Outdoors & 63{,}216 & 24{,}000 & 535{,}010 & 8.5 \\
\hline
\end{tabular}
\caption{Interaction statistics per category after user sampling
and RecBole formatting. Interaction counts include all splits (train/valid/test).}
\label{tab:dataset-stats}
\end{table}

\subsection*{PCF Source Selection}

For each Amazon product, the downstream pipeline assigns a single PCF value before re-ranking. This value is called the \emph{selected} PCF because it is chosen from the available estimator outputs after the held-out Carbon Catalogue comparison. Since the few-shot retrieval-augmented LLM performs best on consumer-scale absolute-error metrics, it is the default selected source. In the reported run, 71{,}989 of 72{,}000 scored products (99.98\%) received a valid
few-shot LLM estimate from \texttt{Qwen/Qwen2.5-3B-Instruct}, so the selected PCF column
is effectively identical to the few-shot prediction across the Amazon catalog. The 11
fallback items are assigned the neighbour-average estimate because the few-shot output was missing or not parseable under the strict numeric parser.

\subsection*{PCF Estimation Diagnostics}

\begin{figure}[H]
  \centering
  \begin{subfigure}[t]{0.48\textwidth}
    \includegraphics[width=\textwidth]{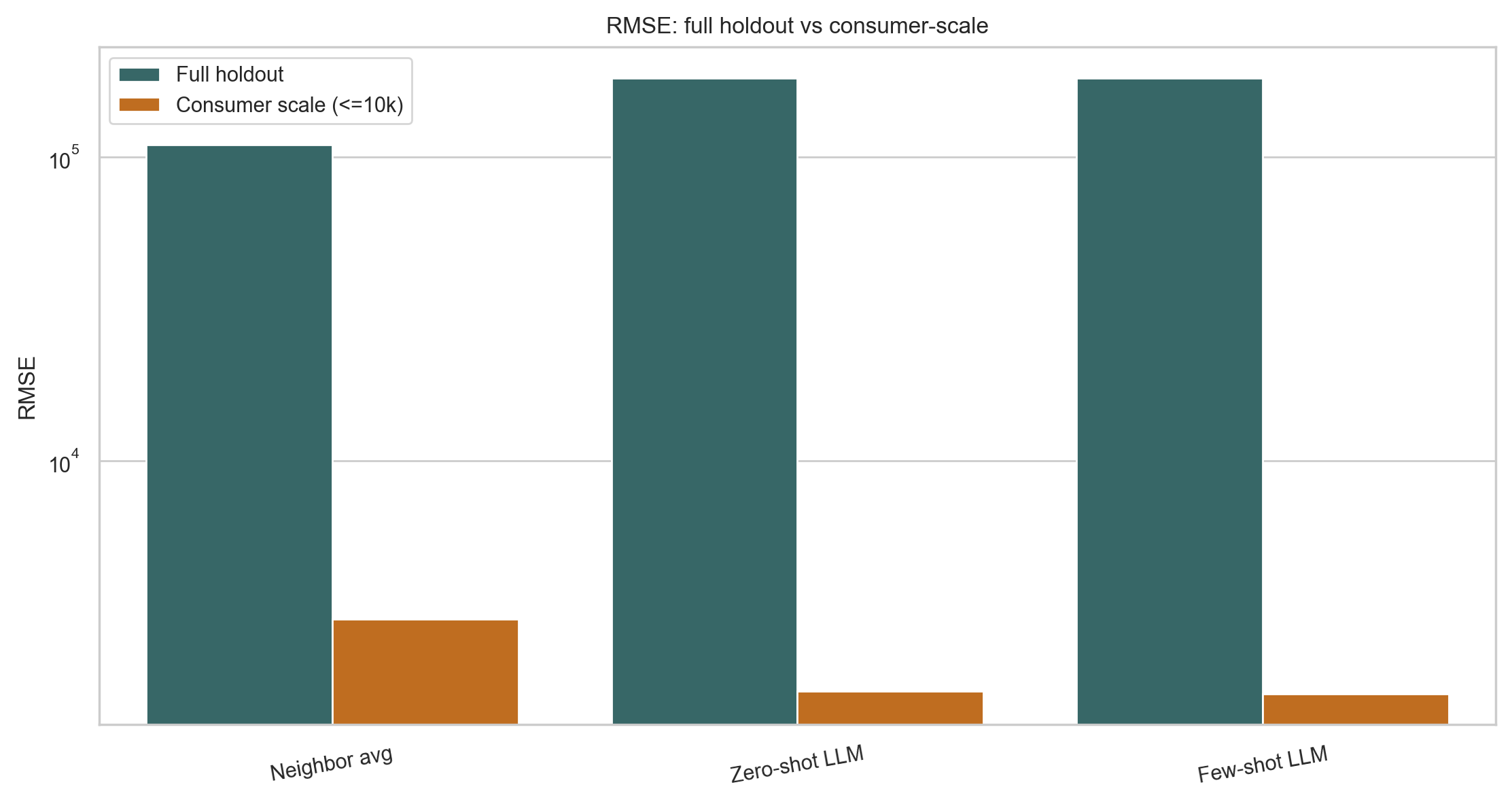}
    \caption{RMSE.}
  \end{subfigure}
  \hfill
  \begin{subfigure}[t]{0.48\textwidth}
    \includegraphics[width=\textwidth]{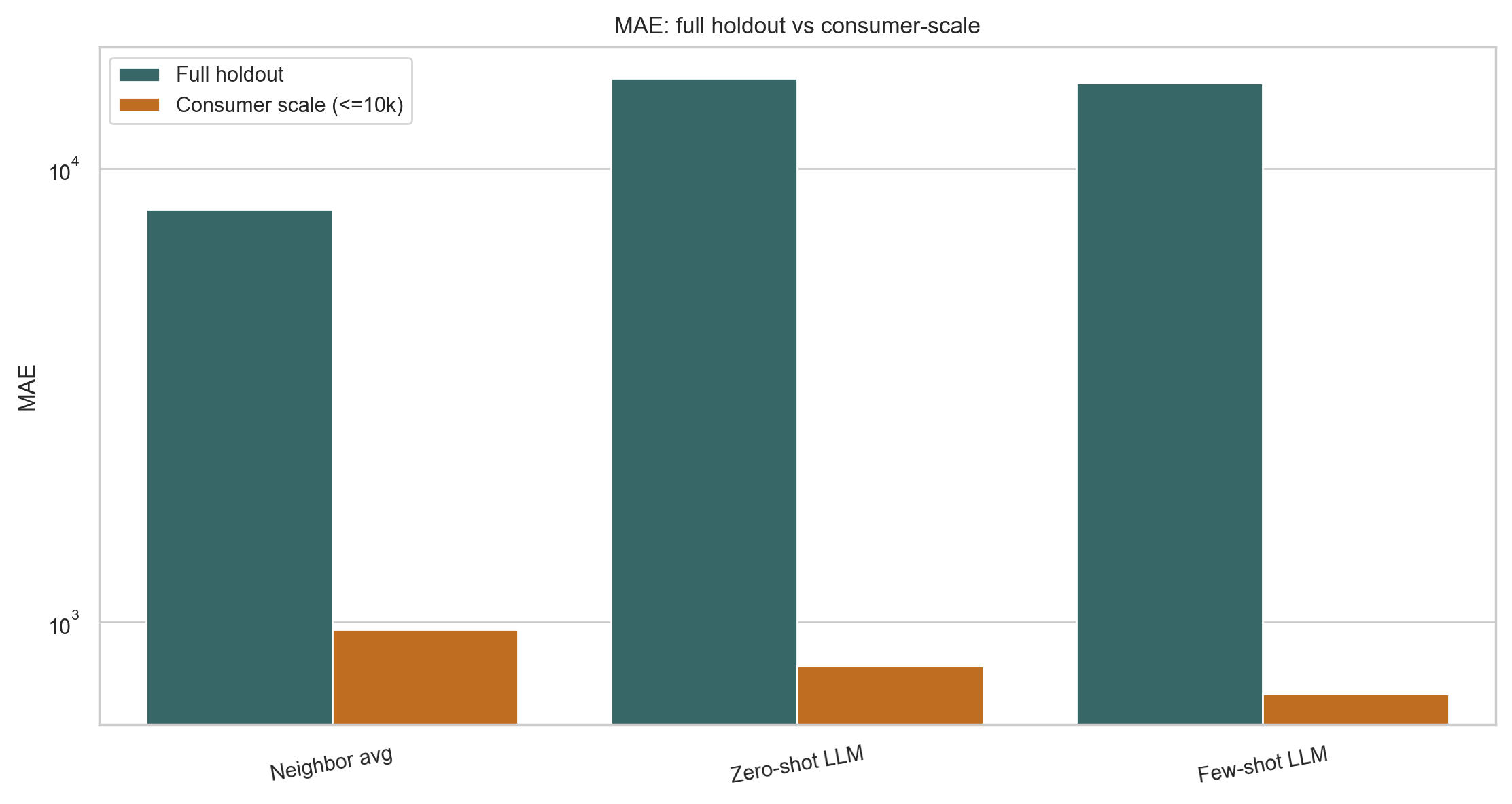}
    \caption{MAE.}
  \end{subfigure}

  \vspace{0.3cm}

  \begin{subfigure}[t]{0.48\textwidth}
    \includegraphics[width=\textwidth]{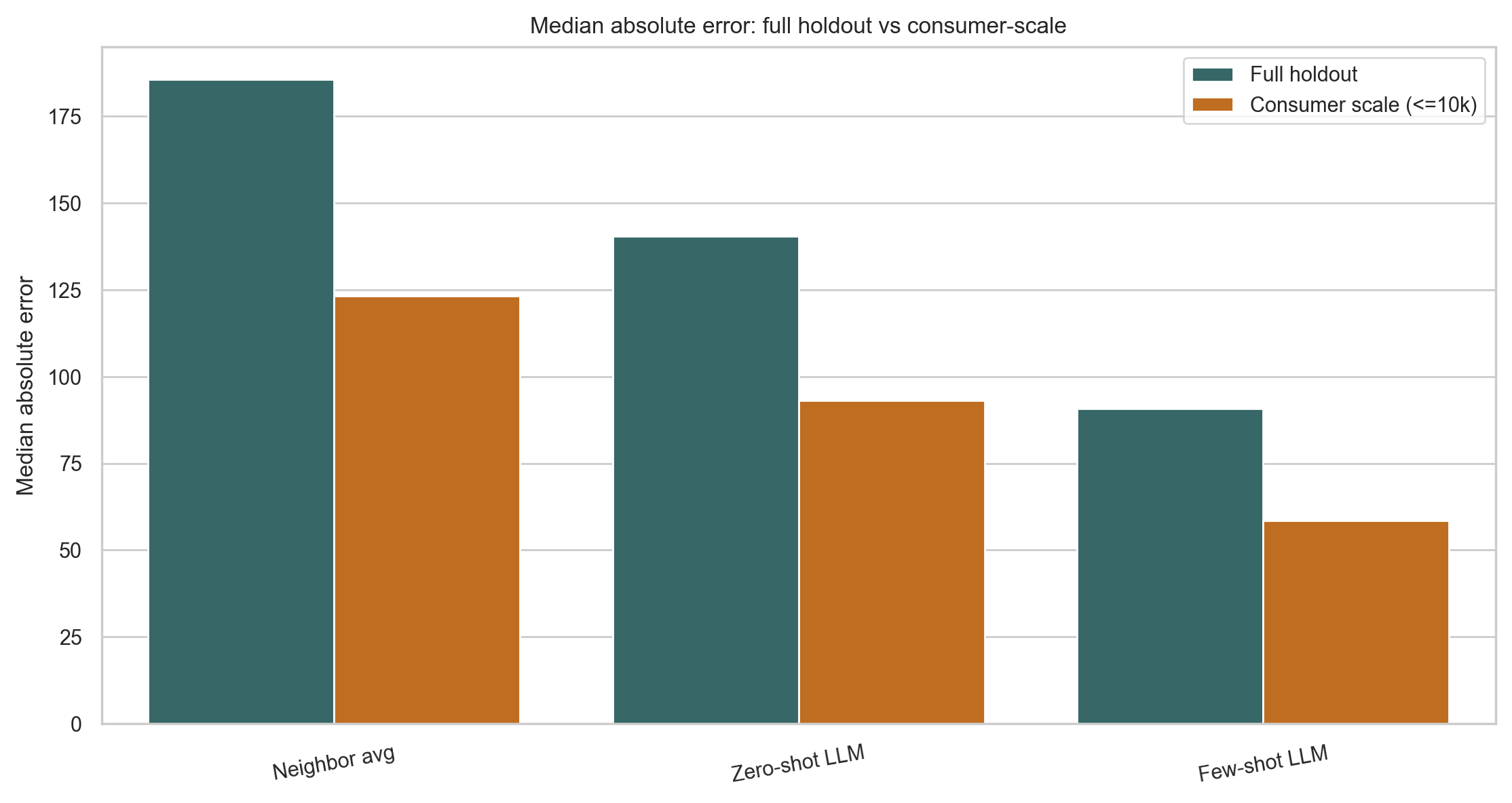}
    \caption{Median absolute error.}
  \end{subfigure}
  \hfill
  \begin{subfigure}[t]{0.48\textwidth}
    \includegraphics[width=\textwidth]{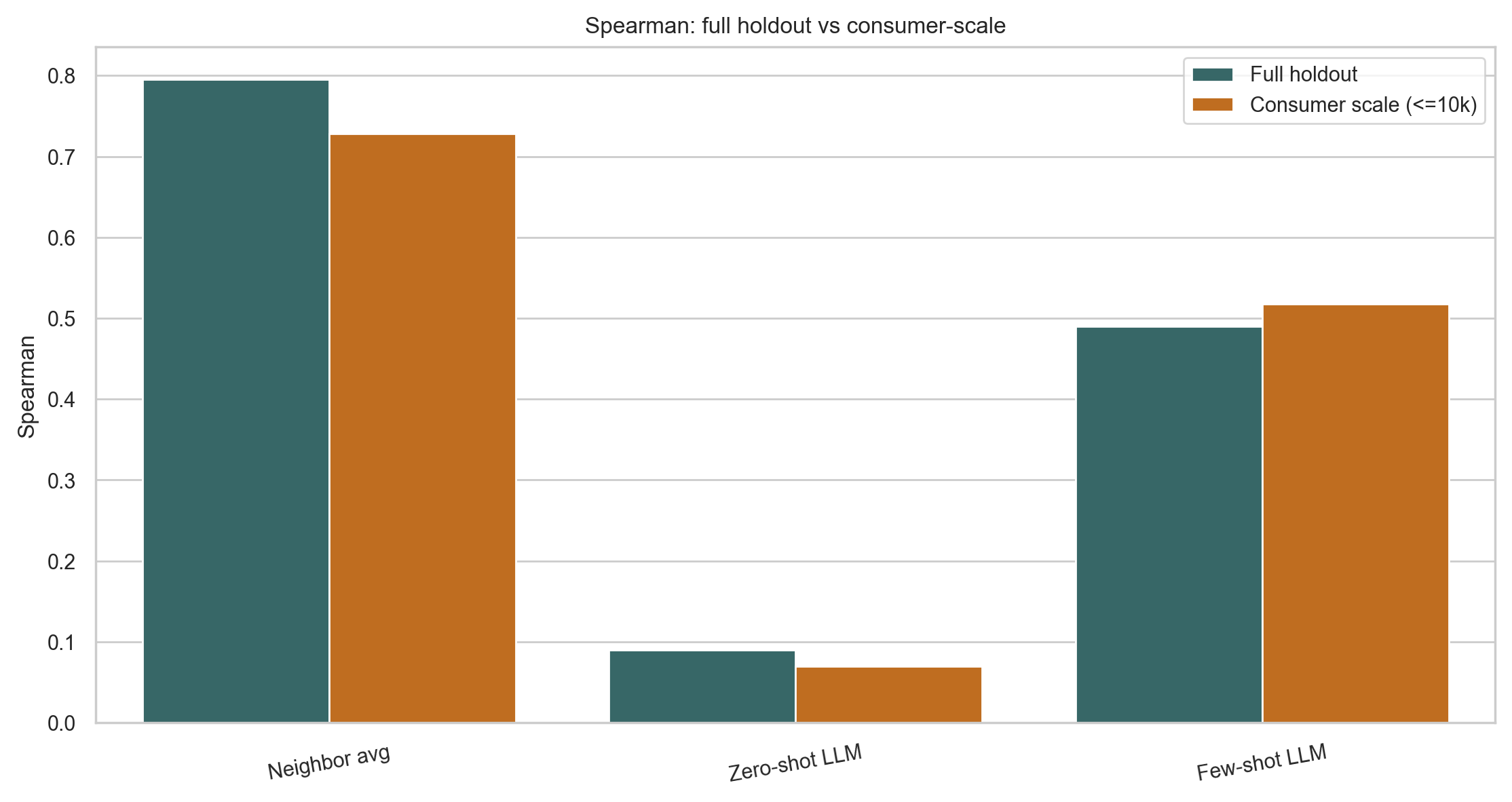}
    \caption{Spearman rank correlation.}
  \end{subfigure}
  \caption{PCF estimation accuracy on the full 866-item holdout and the
    consumer-scale subset ($\mathrm{PCF}_{\mathrm{true}} \leq 10{,}000$).
    The full-holdout error metrics are dominated by a small number of
    extreme industrial products, whereas the consumer-scale subset is
    better aligned with the Amazon recommendation setting studied
    downstream.}
  \label{fig:pcf-subset-accuracy}
\end{figure}

\begin{figure}[H]
  \centering
  \includegraphics[width=0.92\textwidth]{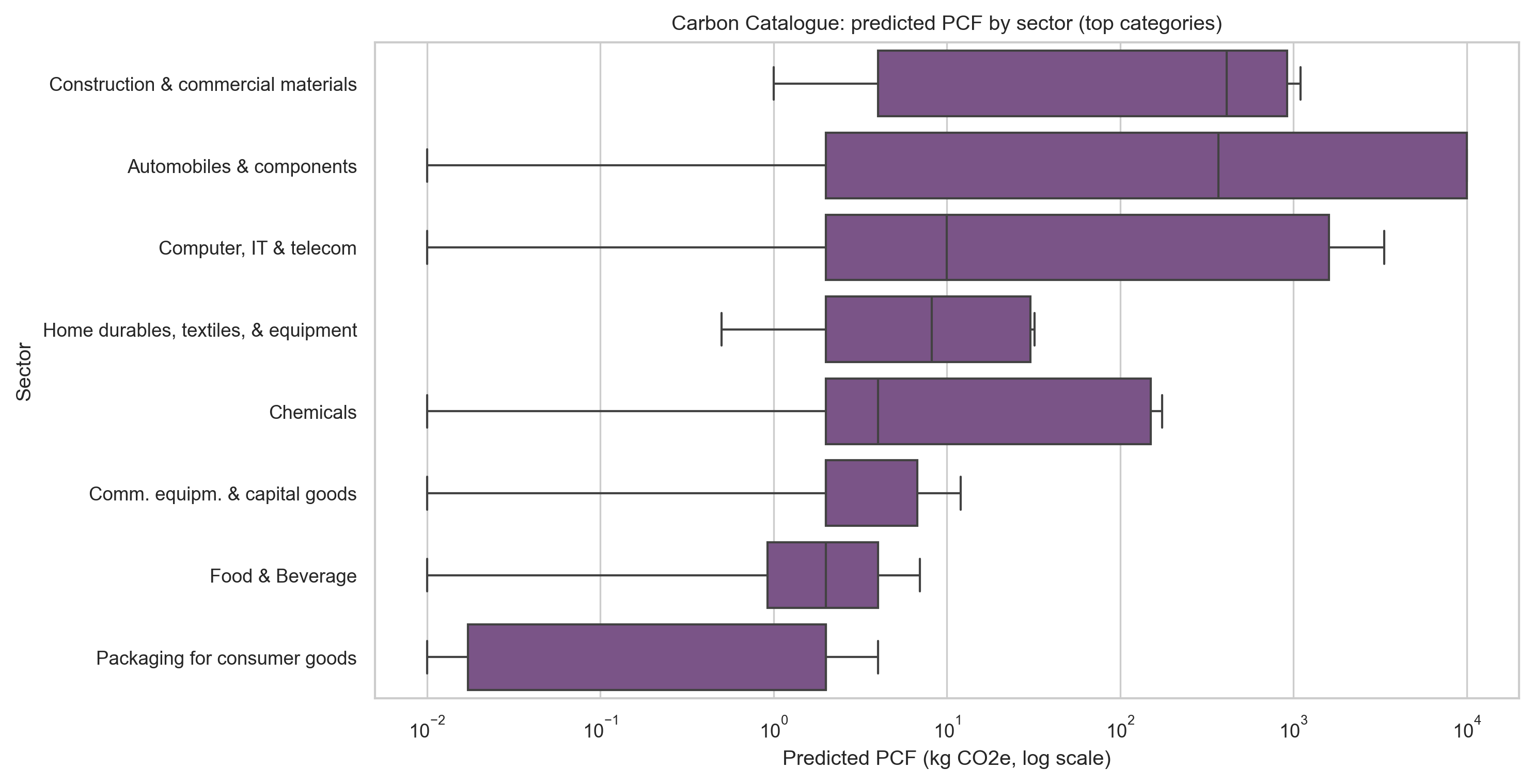}
  \caption{Distribution of predicted PCF (few-shot) across Carbon Catalogue sectors used in PCF estimation
    diagnostics. Sectors span a wide range, including a heavy industrial
    tail that drives large errors on the full 866-item holdout relative to
    the consumer-scale benchmark in Table~\ref{tab:pcf}.}
  \label{fig:pcf-downstream-dist}
\end{figure}

\begin{figure}[H]
  \centering
  \includegraphics[width=0.82\textwidth]{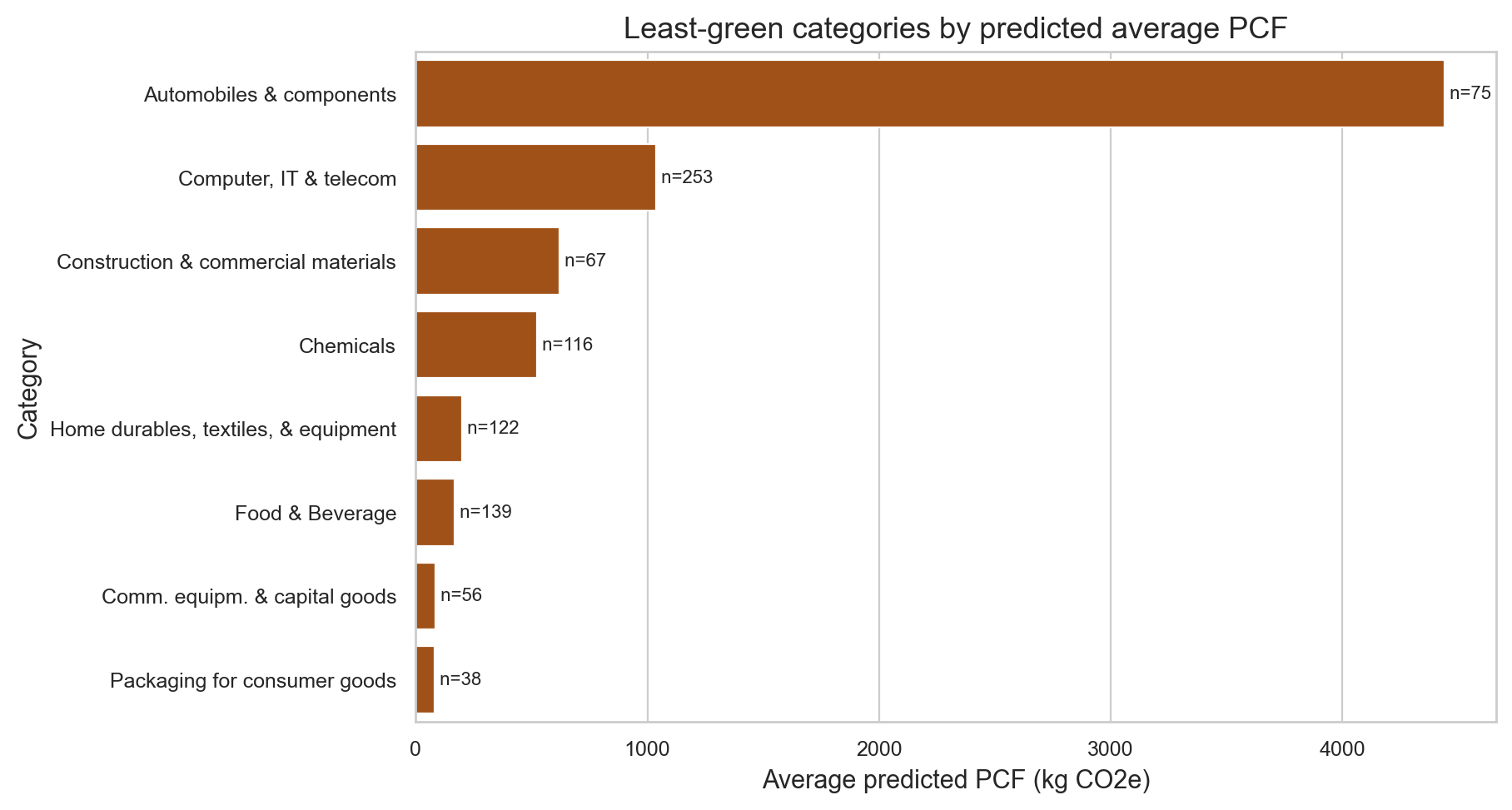}
  \caption{Highest-average-PCF categories in the Carbon Catalogue based on
    the predicted values. A few heavy industrial sectors dominate the
    extreme tail, which is precisely why full-holdout RMSE paints a much
    harsher picture of LLM-based estimation than the consumer-scale subset
    used as the main benchmark.}
  \label{fig:pcf-sector-tail}
\end{figure}

\subsection*{Additional Results Figures}

\begin{figure}[H]
  \centering
  \begin{subfigure}[t]{0.32\textwidth}
    \includegraphics[width=\textwidth]{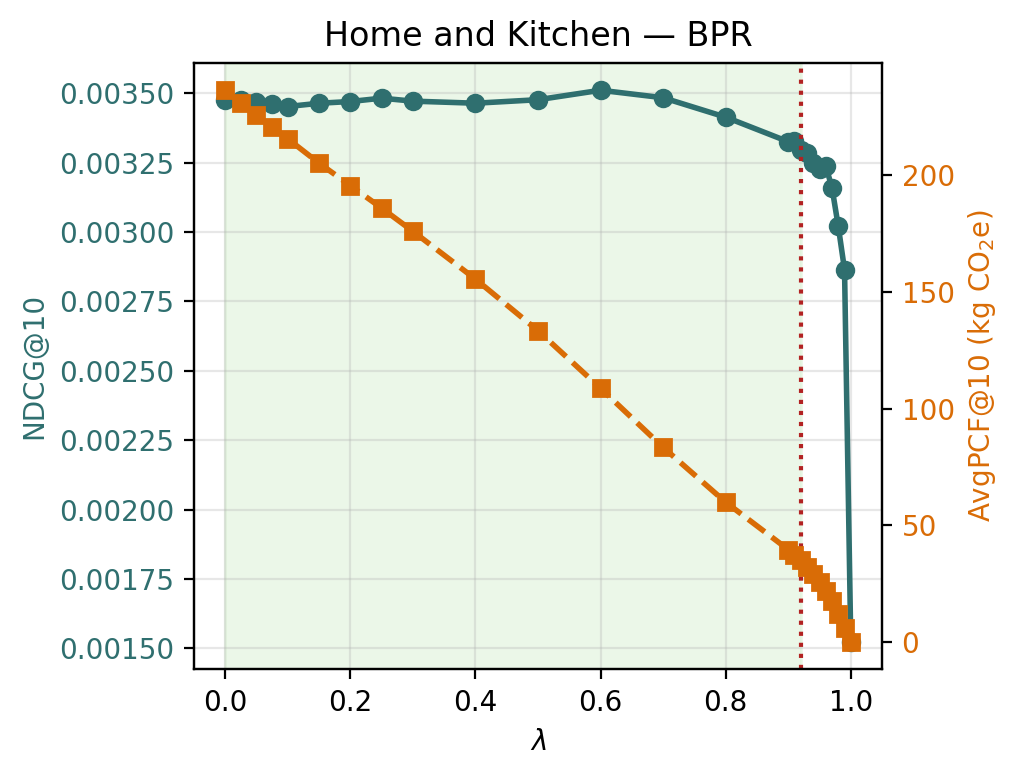}
    \caption{BPR.}
  \end{subfigure}
  \hfill
  \begin{subfigure}[t]{0.32\textwidth}
    \includegraphics[width=\textwidth]{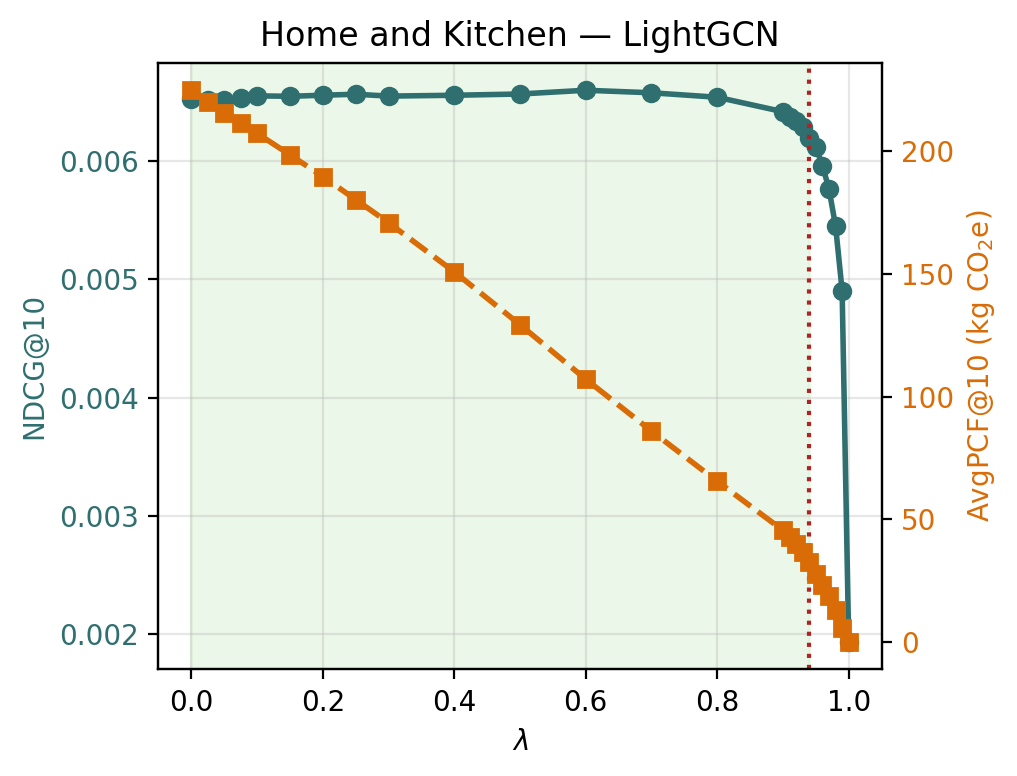}
    \caption{LightGCN.}
  \end{subfigure}
  \hfill
  \begin{subfigure}[t]{0.32\textwidth}
    \includegraphics[width=\textwidth]{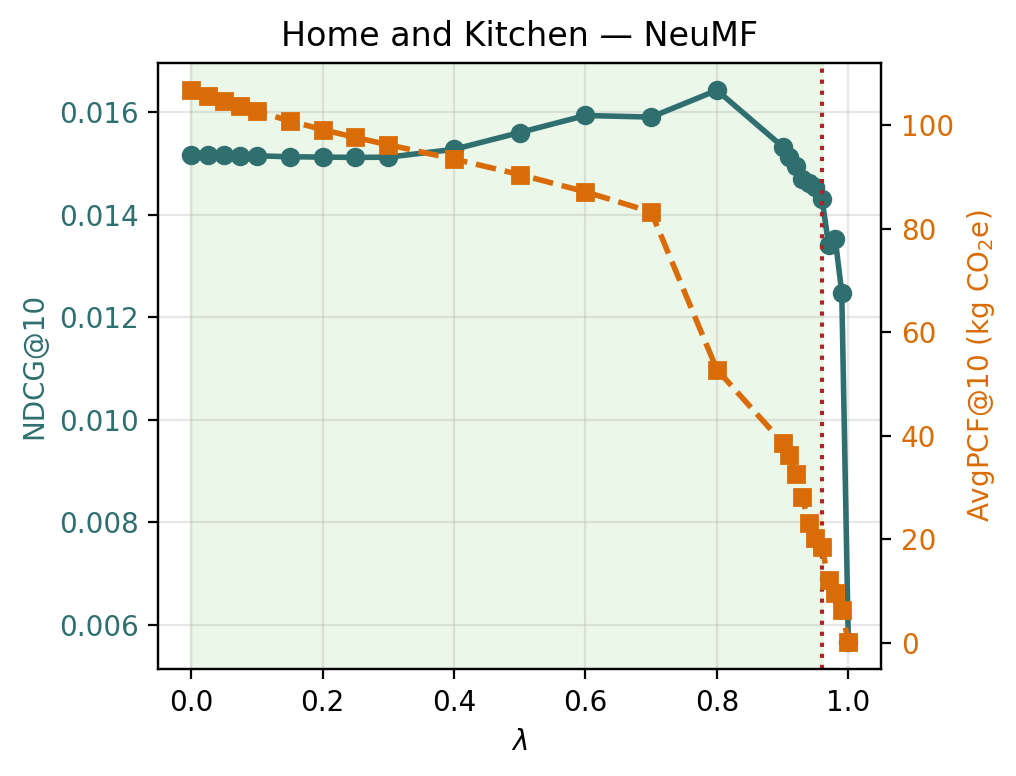}
    \caption{NeuMF.}
  \end{subfigure}
  \caption{$\lambda$ sensitivity for the Home and Kitchen category, shown
    in the same format as Figure~\ref{fig:lambda-electronics}; the shaded green region marks settings within a 5\% NDCG@10 drop from the $\lambda=0$ baseline.}
  \label{fig:lambda-home}
\end{figure}

\begin{figure}[H]
  \centering
  \begin{subfigure}[t]{0.32\textwidth}
    \includegraphics[width=\textwidth]{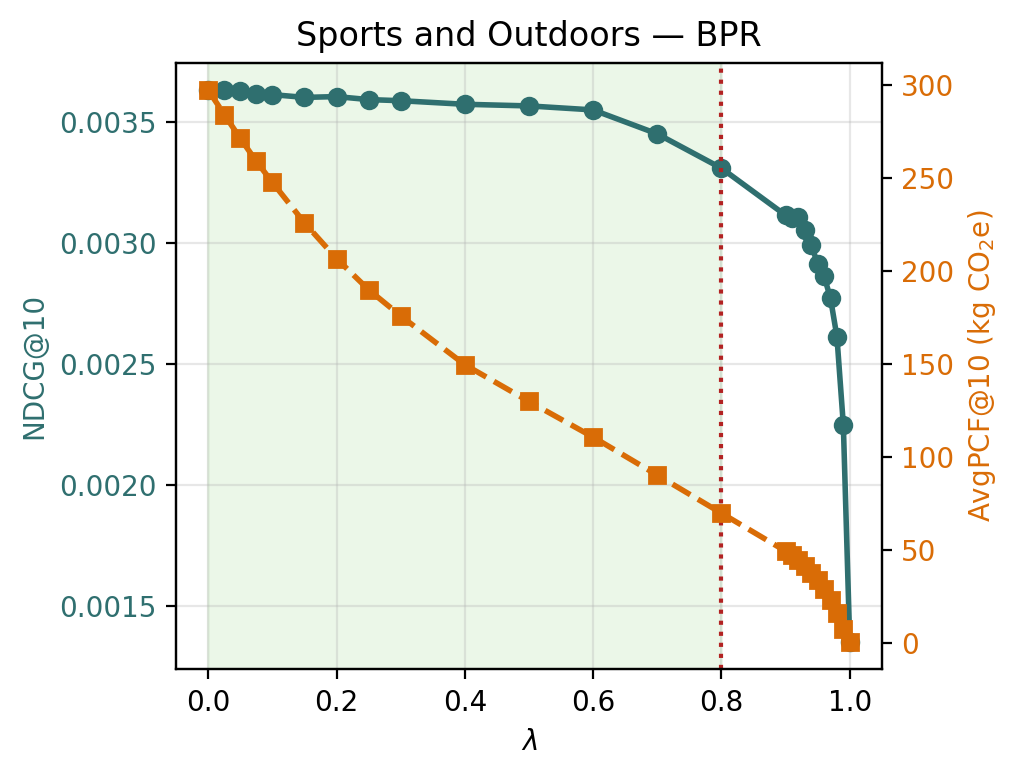}
    \caption{BPR.}
  \end{subfigure}
  \hfill
  \begin{subfigure}[t]{0.32\textwidth}
    \includegraphics[width=\textwidth]{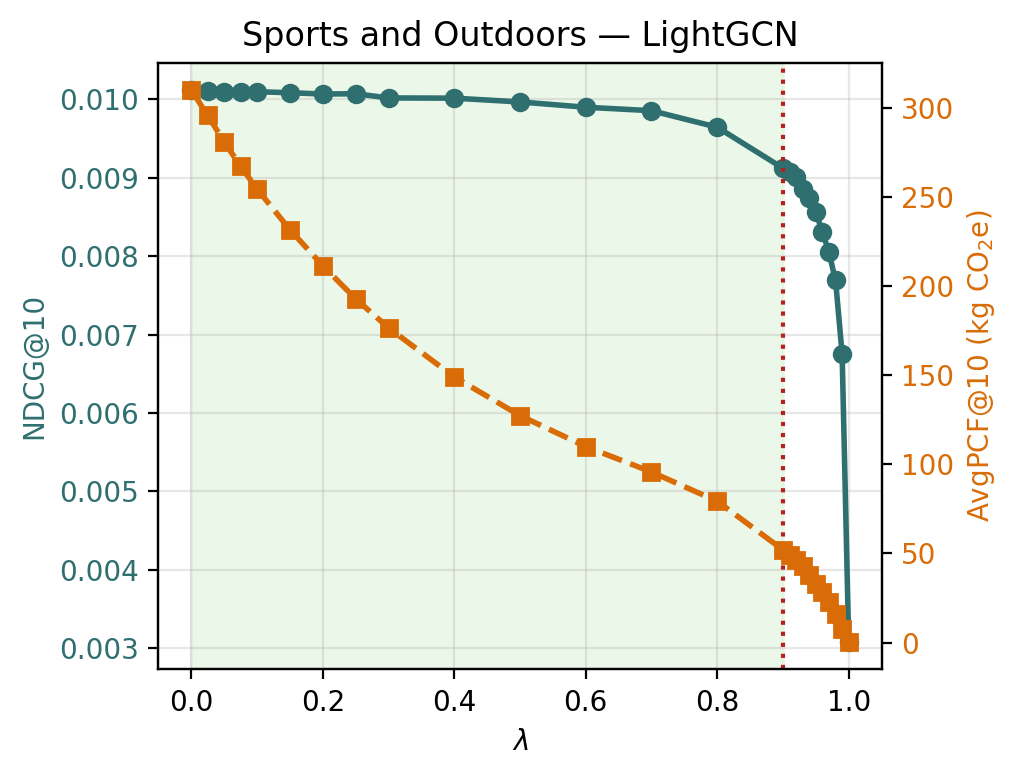}
    \caption{LightGCN.}
  \end{subfigure}
  \hfill
  \begin{subfigure}[t]{0.32\textwidth}
    \includegraphics[width=\textwidth]{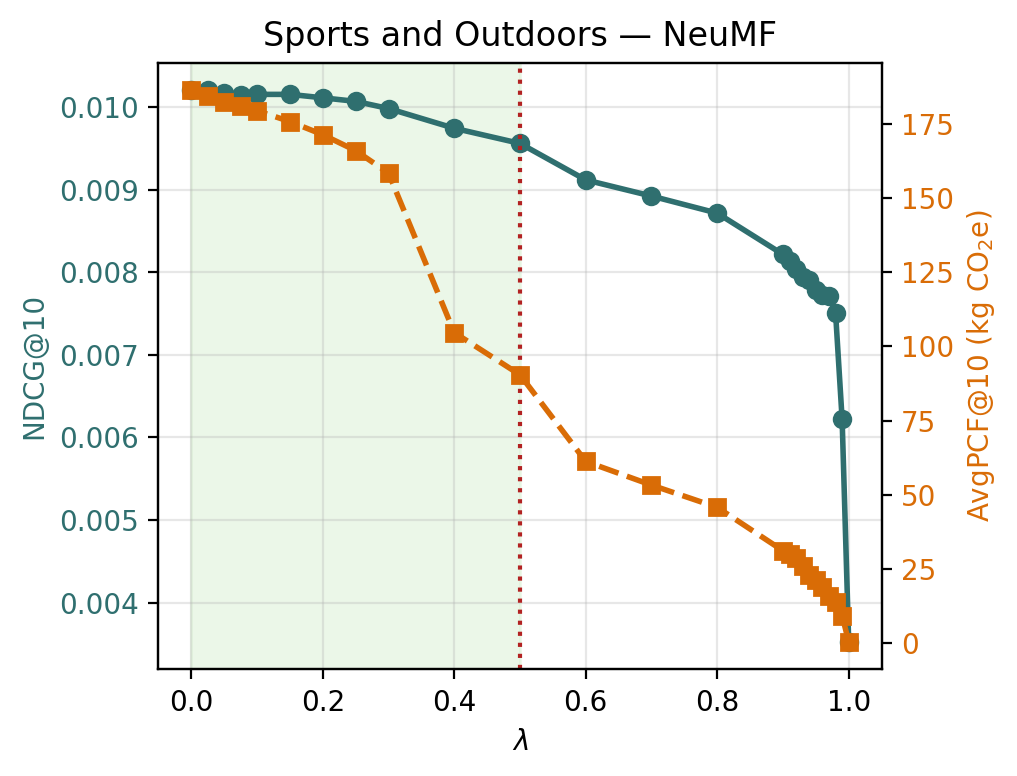}
    \caption{NeuMF.}
  \end{subfigure}
  \caption{$\lambda$ sensitivity for the Sports and Outdoors category,
    shown in the same format as Figure~\ref{fig:lambda-electronics}; the shaded green region marks settings within a 5\% NDCG@10 drop from the $\lambda=0$ baseline.}
  \label{fig:lambda-sports}
\end{figure}

\begin{figure}[H]
  \centering
  \begin{subfigure}[t]{0.32\textwidth}
    \includegraphics[width=\textwidth]{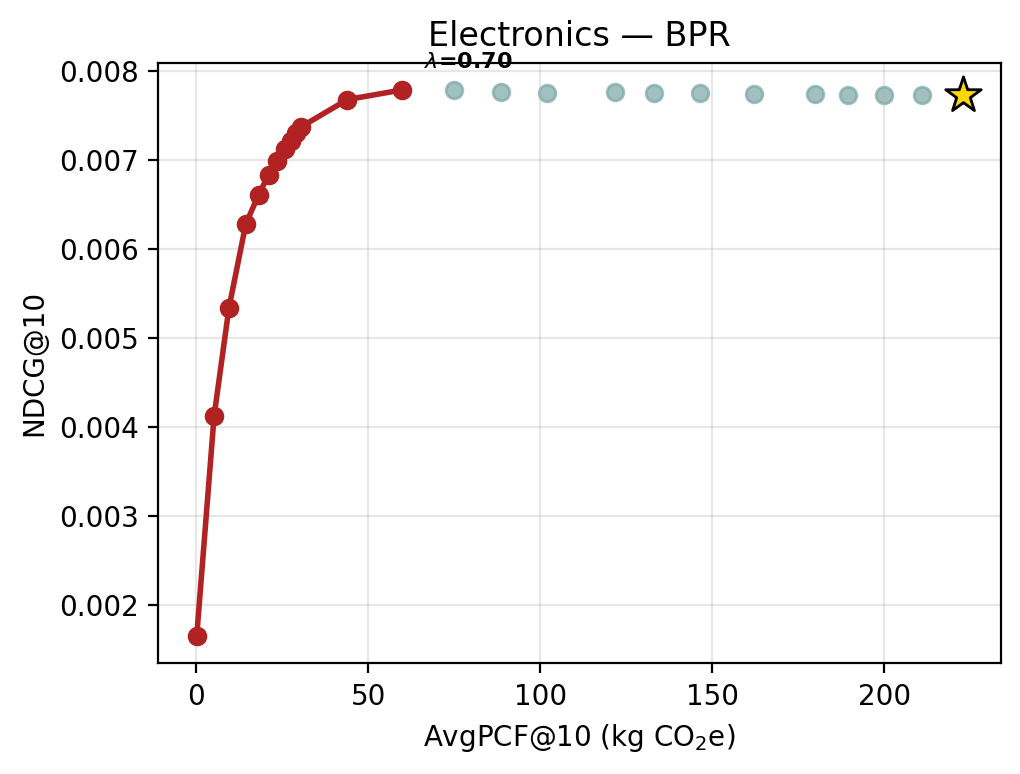}
    \caption{BPR.}
  \end{subfigure}
  \hfill
  \begin{subfigure}[t]{0.32\textwidth}
    \includegraphics[width=\textwidth]{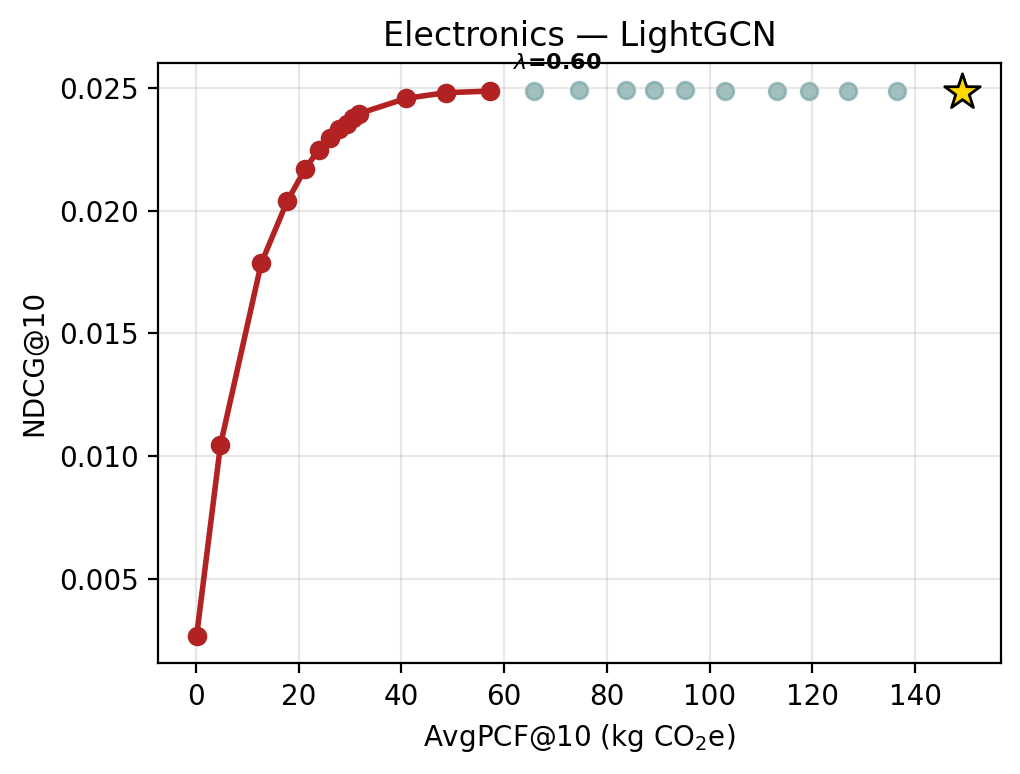}
    \caption{LightGCN.}
  \end{subfigure}
  \hfill
  \begin{subfigure}[t]{0.32\textwidth}
    \includegraphics[width=\textwidth]{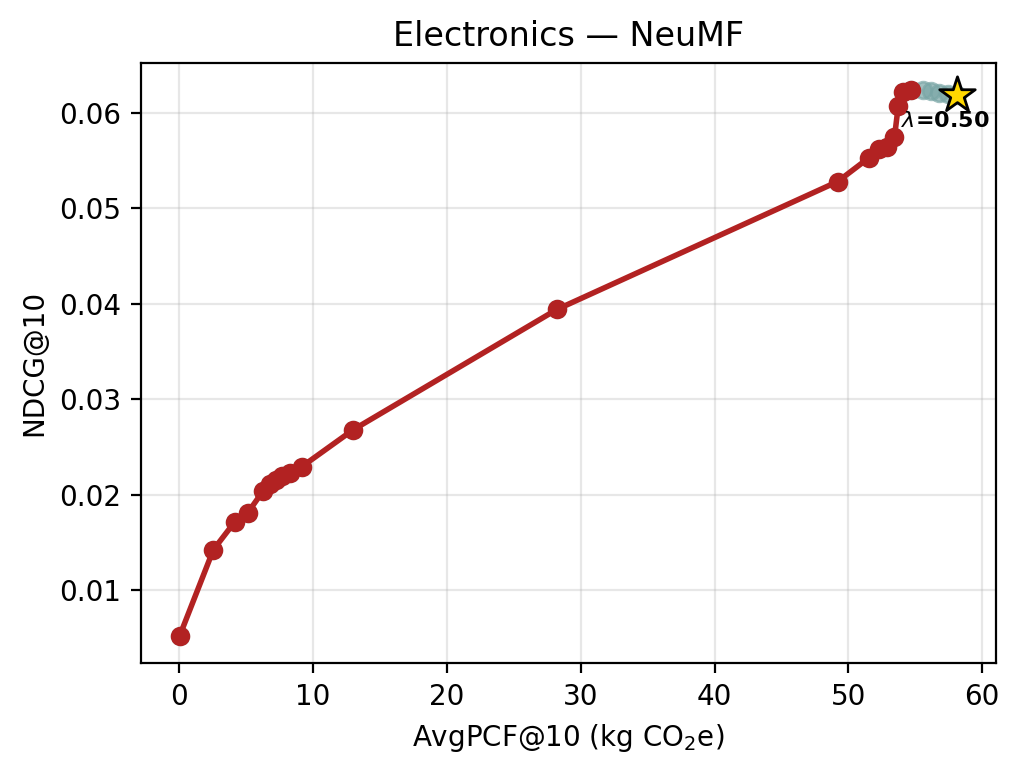}
    \caption{NeuMF.}
  \end{subfigure}
  \caption{Engagement and carbon Pareto frontiers for Electronics. Each point
    corresponds to one $\lambda$ value; faint gray points are evaluated but dominated settings, red points and lines denote
    Pareto-optimal operating points, and the star marks the
    engagement-only baseline ($\lambda=0$).}
  \label{fig:pareto-electronics}
\end{figure}

\begin{figure}[H]
  \centering
  \begin{subfigure}[t]{0.32\textwidth}
    \includegraphics[width=\textwidth]{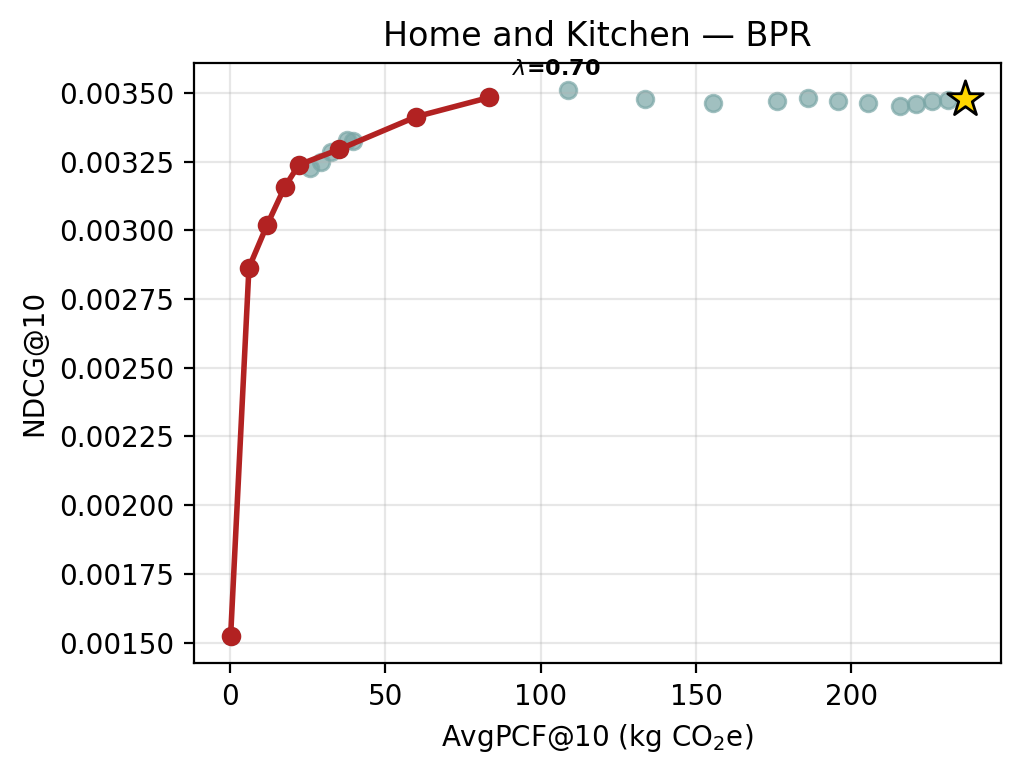}
    \caption{BPR.}
  \end{subfigure}
  \hfill
  \begin{subfigure}[t]{0.32\textwidth}
    \includegraphics[width=\textwidth]{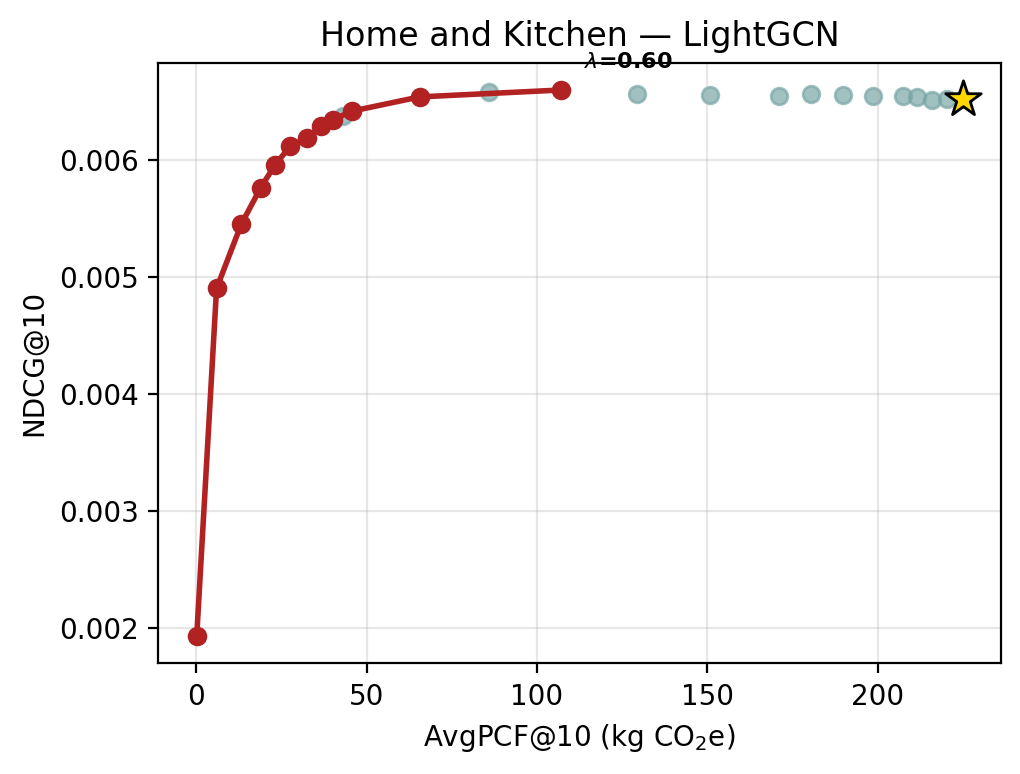}
    \caption{LightGCN.}
  \end{subfigure}
  \hfill
  \begin{subfigure}[t]{0.32\textwidth}
    \includegraphics[width=\textwidth]{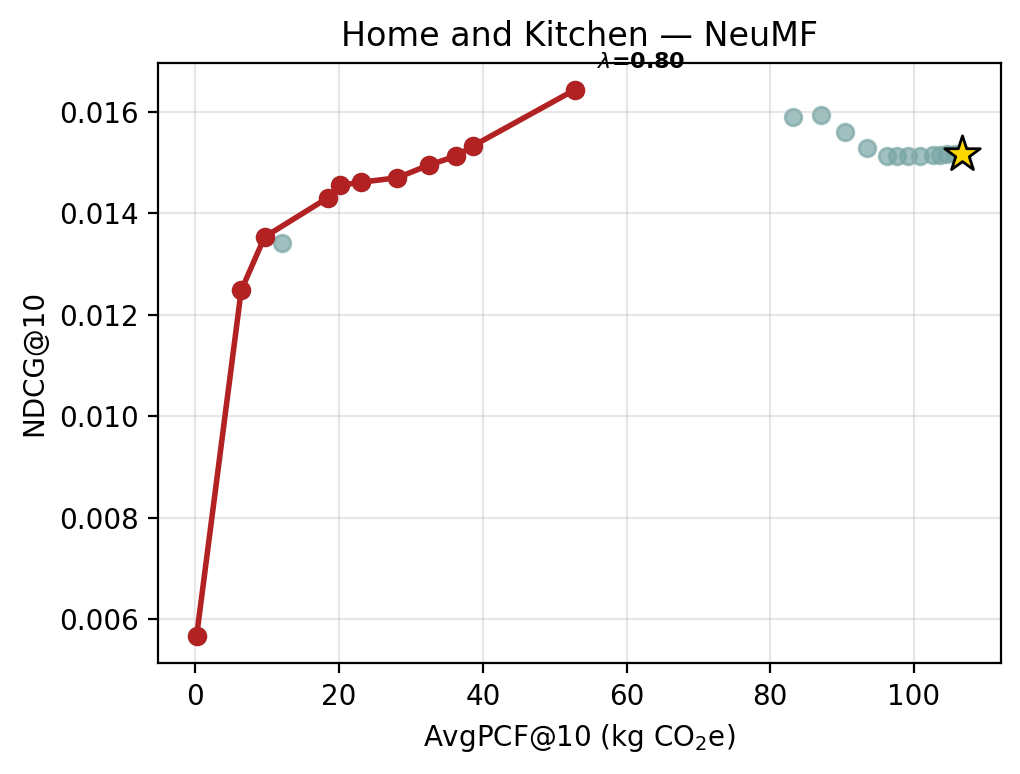}
    \caption{NeuMF.}
  \end{subfigure}
  \caption{Engagement and carbon Pareto frontiers for Home and Kitchen,
    shown in the same format as Figure~\ref{fig:pareto-electronics}, with faint gray points denoting dominated $\lambda$ settings.}
  \label{fig:pareto-home}
\end{figure}

\begin{figure}[H]
  \centering
  \begin{subfigure}[t]{0.32\textwidth}
    \includegraphics[width=\textwidth]{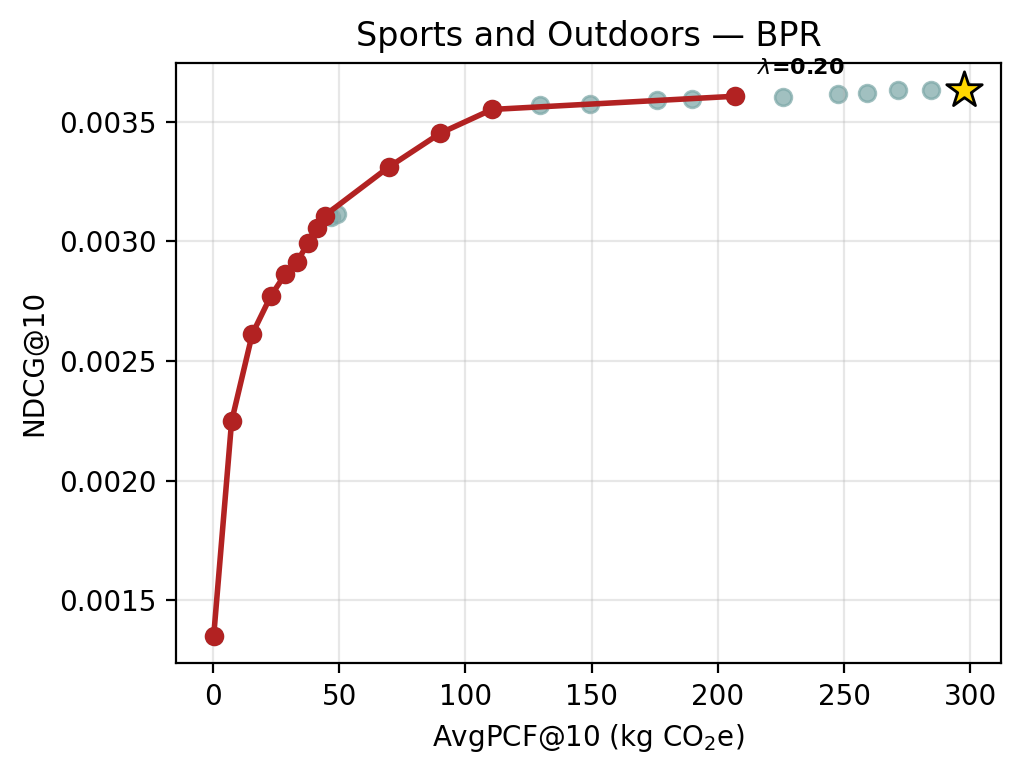}
    \caption{BPR.}
  \end{subfigure}
  \hfill
  \begin{subfigure}[t]{0.32\textwidth}
    \includegraphics[width=\textwidth]{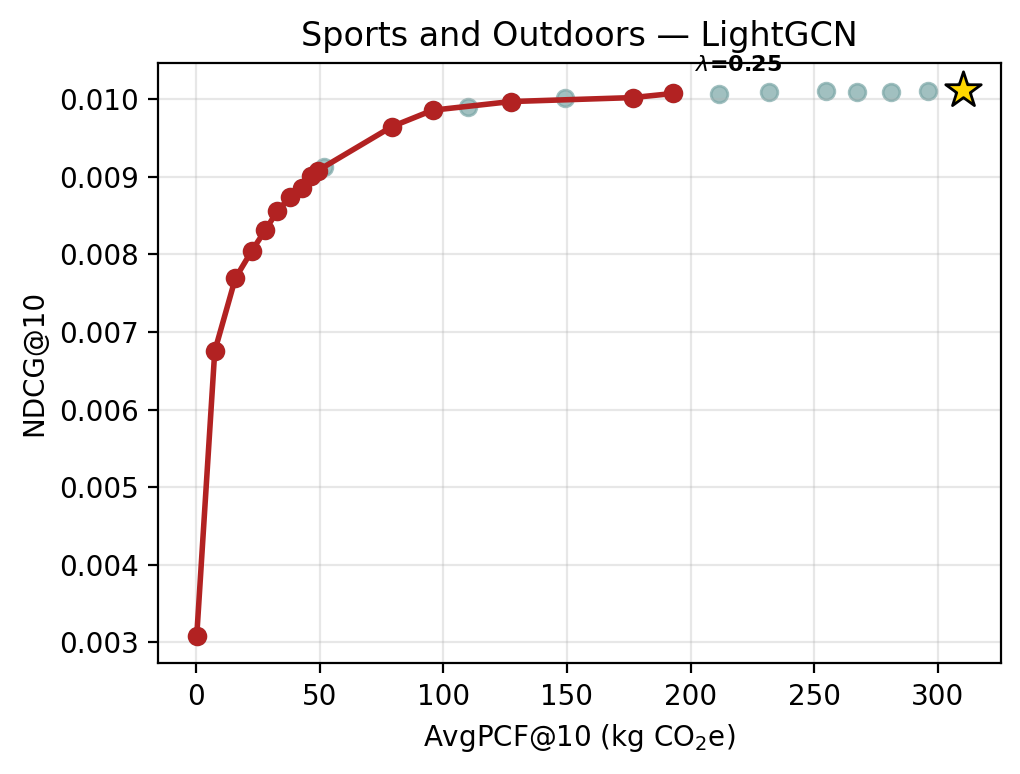}
    \caption{LightGCN.}
  \end{subfigure}
  \hfill
  \begin{subfigure}[t]{0.32\textwidth}
    \includegraphics[width=\textwidth]{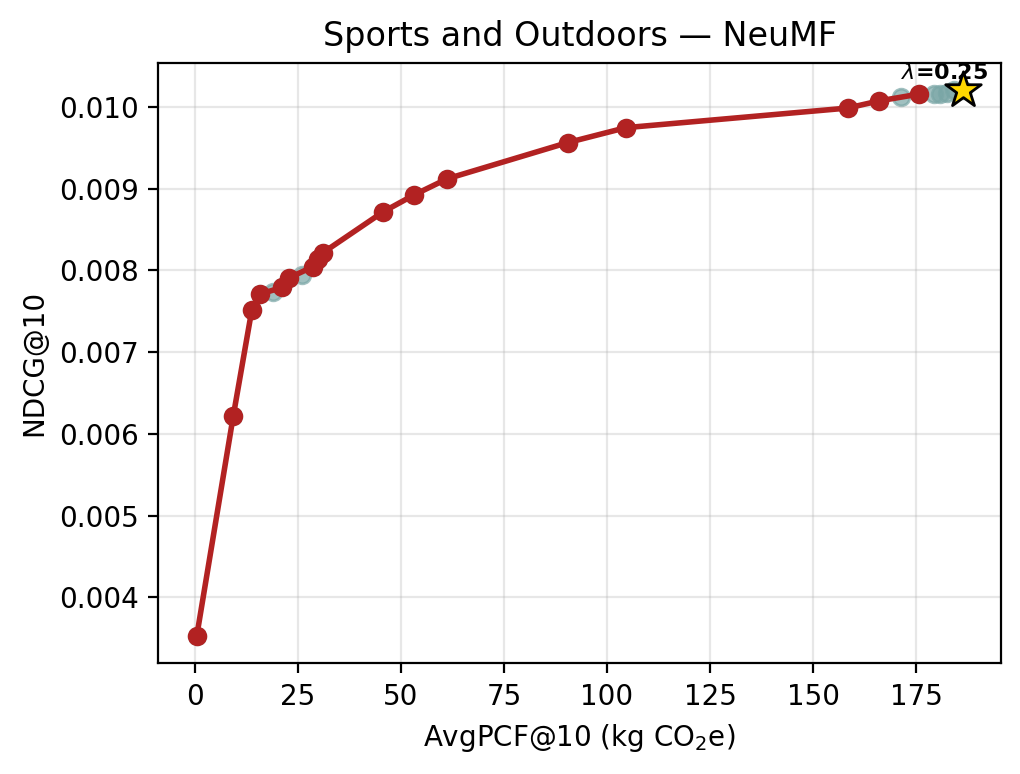}
    \caption{NeuMF.}
  \end{subfigure}
  \caption{Engagement and carbon Pareto frontiers for Sports and Outdoors,
    shown in the same format as Figure~\ref{fig:pareto-electronics}, with faint gray points denoting dominated $\lambda$ settings.}
  \label{fig:pareto-sports}
\end{figure}

\begin{figure}[H]
  \centering
  \includegraphics[width=\textwidth]{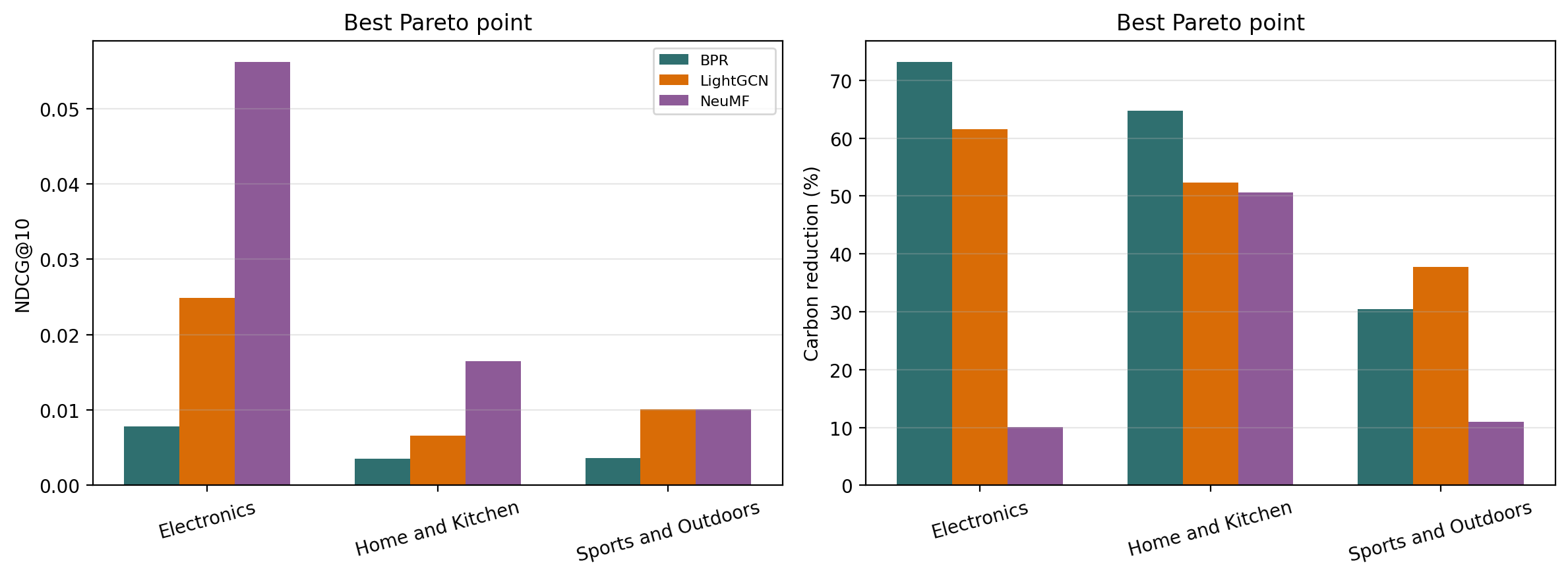}
  \caption{Best Pareto operating point per model and category, subject to
    at least 10\% carbon reduction relative to $\lambda=0$. The left
    panel reports NDCG@10 at the chosen operating point; the right panel
    reports the corresponding carbon reduction.}
  \label{fig:best-pareto-summary}
\end{figure}

\begin{figure}[H]
  \centering
  \begin{subfigure}[t]{0.32\textwidth}
    \includegraphics[width=\textwidth]{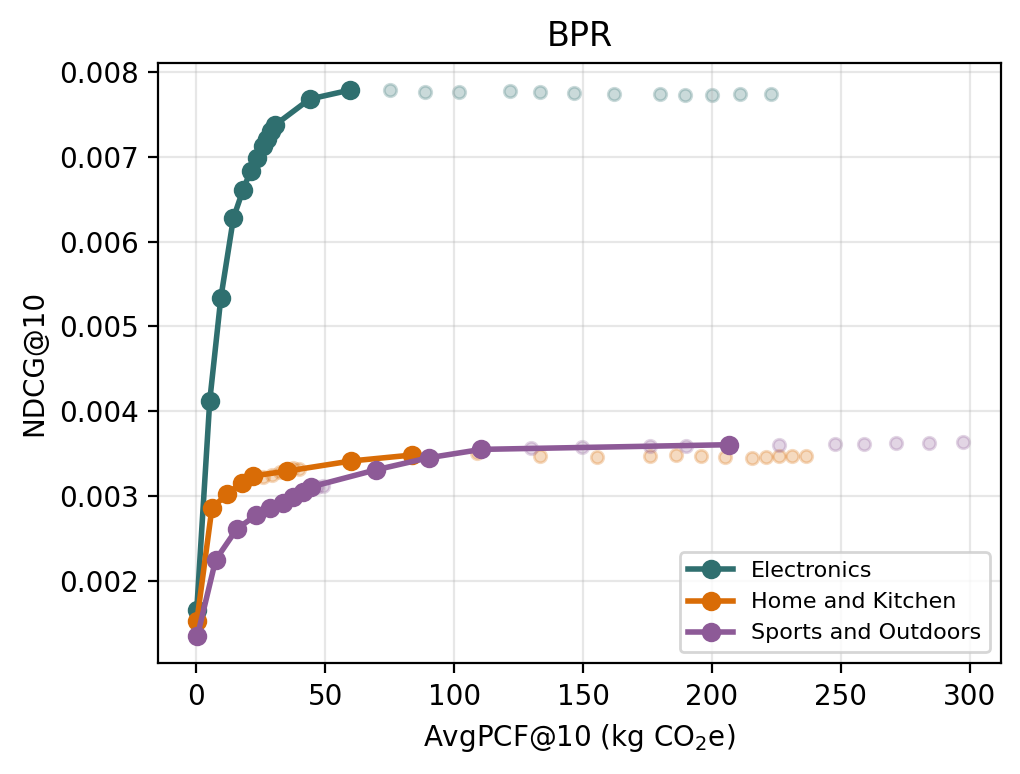}
    \caption{BPR.}
  \end{subfigure}
  \hfill
  \begin{subfigure}[t]{0.32\textwidth}
    \includegraphics[width=\textwidth]{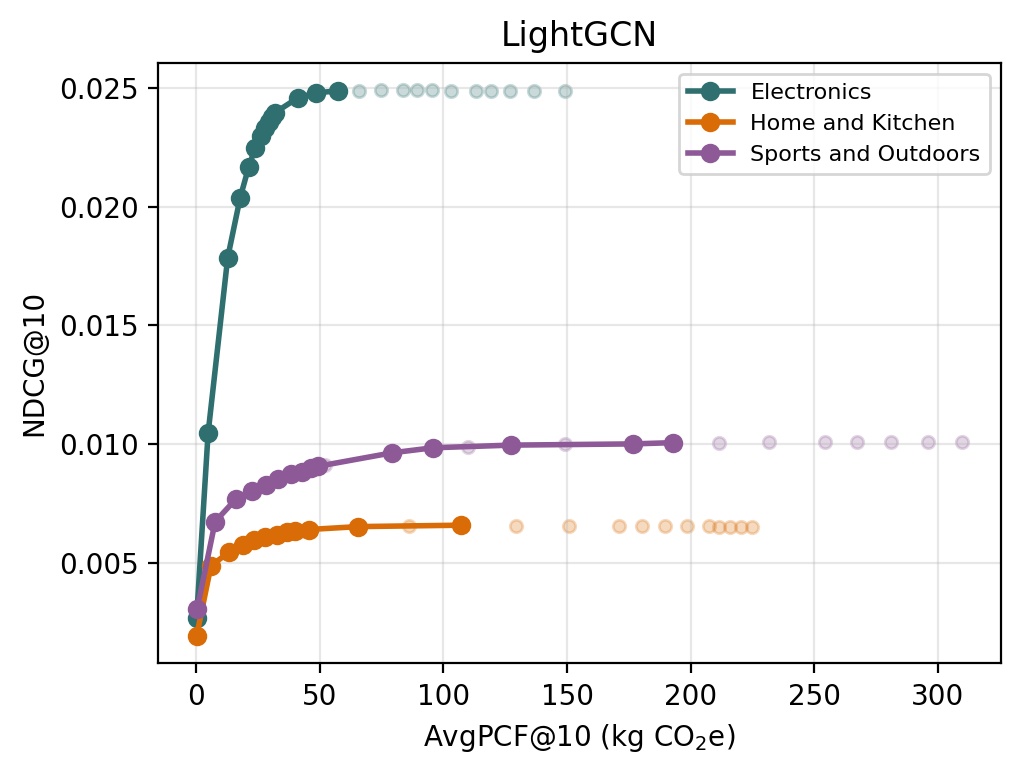}
    \caption{LightGCN.}
  \end{subfigure}
  \hfill
  \begin{subfigure}[t]{0.32\textwidth}
    \includegraphics[width=\textwidth]{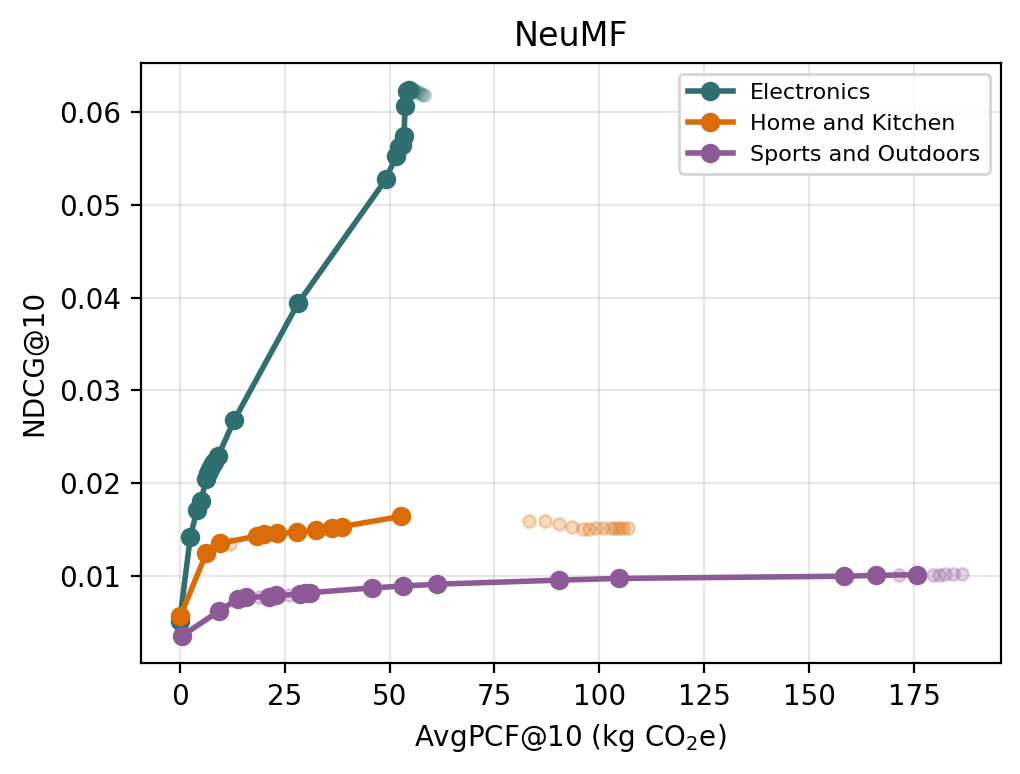}
    \caption{NeuMF.}
  \end{subfigure}
  \caption{Cross-category Pareto frontier comparison, one panel per model.
    Each panel overlays the three product categories to isolate category
    effects while holding the recommendation backbone fixed.}
  \label{fig:cross-category-pareto}
\end{figure}

\subsection*{LLM Prompt Template}

The few-shot prompt template used for PCF estimation is shown in
Figure~\ref{fig:llm-prompt-appendix}. The five nearest Carbon Catalogue neighbours
retrieved by cosine similarity are inserted as labeled in-context examples before the
target Amazon product.

\begin{figure}[H]
  \centering
  \scalebox{0.75}{\definecolor{promptbg}     {HTML}{FAFAF8}
\definecolor{promptborder} {HTML}{C8C6BC}
\definecolor{headerbg}     {HTML}{EEEDFE}
\definecolor{headerborder} {HTML}{534AB7}
\definecolor{exbg}         {HTML}{E1F5EE}
\definecolor{exborder}     {HTML}{0F6E56}
\definecolor{querybg}      {HTML}{FAEEDA}
\definecolor{queryborder}  {HTML}{854F0B}
\definecolor{rolecolor}    {HTML}{3C3489}
\definecolor{commentcolor} {HTML}{888780}
\definecolor{excolor}      {HTML}{085041}
\definecolor{querycolor}   {HTML}{633806}
\definecolor{keycolor}     {HTML}{185FA5}
\definecolor{valcolor}     {HTML}{2C2C2A}
\definecolor{divcolor}     {HTML}{D3D1C7}

\newcommand{\sectionpill}[4]{%
  \tikz[baseline=-0.6ex]\node[
    rectangle, rounded corners=3pt,
    fill=#1, draw=#2, line width=0.4pt,
    inner xsep=5pt, inner ysep=2pt,
    font=\scriptsize\bfseries, text=#4
  ] {#3};%
}

\newlength{\promptwidth}
\setlength{\promptwidth}{13.0cm}

\newlength{\bodywidth}
\setlength{\bodywidth}{12.32cm}

\newcommand{\promptrule}{%
  \textcolor{divcolor}{\rule{\bodywidth}{0.3pt}}%
}

\begin{tikzpicture}
\node[
  rectangle,
  rounded corners=8pt,
  fill=promptbg,
  draw=promptborder,
  line width=0.6pt,
  inner sep=0pt,
  text width=\promptwidth,
  align=left
] (prompt) {%

  \noindent\colorbox{headerbg}{%
    \hspace{12pt}%
    \begin{minipage}[c]{\dimexpr\promptwidth-24pt\relax}%
      \vspace{6pt}%
      {\small\textbf{\textcolor{headerborder}{PCF estimation prompt}}%
      \quad\textcolor{commentcolor}{few-shot $\cdot$ chain-of-thought $\cdot$ structured output}}%
      \vspace{6pt}%
    \end{minipage}%
    \hspace{0pt}%
  }%

  \hspace{12pt}%
  \begin{minipage}[t]{\bodywidth}
  \vspace{8pt}

  \sectionpill{headerbg}{headerborder}{System}{rolecolor}\\[4pt]
  {\small\ttfamily\textcolor{rolecolor}{%
    You are an expert in product life-cycle assessment (LCA) and
    carbon footprint estimation. Given a product description, estimate
    its Product Carbon Footprint (PCF) in kilograms of CO\textsubscript{2}
    equivalent (kg\,CO\textsubscript{2}e), covering the full product life cycle
    (manufacturing, transport, use, and end-of-life).\\[2pt]
    Always reason step by step before stating your final answer.\\
    Output your final answer strictly as: \textbf{PCF: X.X\,kg CO\textsubscript{2}e}
  }}
  \vspace{5pt}\\
  \promptrule\\
  \vspace{3pt}

  \sectionpill{headerbg}{headerborder}{Instructions}{rolecolor}\\[4pt]
  {\small\ttfamily\textcolor{valcolor}{%
    Five reference products are provided below, ranked by cosine
    similarity of their sentence embeddings to the query product.
    Use them as calibration anchors. When reasoning, consider:
    material composition, product weight, manufacturing complexity,
    and product category.%
  }}
  \vspace{5pt}\\
  \promptrule\\
  \vspace{3pt}

  \sectionpill{exbg}{exborder}{Reference examples --- ranked by embedding similarity}{excolor}\\[4pt]

  {\small\ttfamily%
    \textcolor{commentcolor}{-- Example 1 (sim: 0.94) --}\\[1pt]
    \textcolor{keycolor}{Product:}\ \ \textcolor{valcolor}{Stainless steel vacuum thermos flask, 500\,ml, double-wall}\\
    \textcolor{keycolor}{Category:}\ \textcolor{valcolor}{Kitchenware}\\
    \textcolor{keycolor}{PCF:}\ \ \ \ \ \textcolor{excolor}{\textbf{4.1\,kg CO\textsubscript{2}e}}
  }\\[2pt]
  {\small\ttfamily%
    \textcolor{commentcolor}{-- Example 2 (sim: 0.91) --}\\[1pt]
    \textcolor{keycolor}{Product:}\ \ \textcolor{valcolor}{Stainless steel coffee tumbler with lid, 400\,ml}\\
    \textcolor{keycolor}{Category:}\ \textcolor{valcolor}{Kitchenware}\\
    \textcolor{keycolor}{PCF:}\ \ \ \ \ \textcolor{excolor}{\textbf{2.8\,kg CO\textsubscript{2}e}}
  }\\[2pt]
  {\small\ttfamily%
    \textcolor{commentcolor}{-- Example 3 (sim: 0.88) --}\\[1pt]
    \textcolor{keycolor}{Product:}\ \ \textcolor{valcolor}{Insulated aluminium water bottle, 600\,ml, sport cap}\\
    \textcolor{keycolor}{Category:}\ \textcolor{valcolor}{Sporting goods}\\
    \textcolor{keycolor}{PCF:}\ \ \ \ \ \textcolor{excolor}{\textbf{3.7\,kg CO\textsubscript{2}e}}
  }\\[2pt]
  {\small\ttfamily%
    \textcolor{commentcolor}{-- Example 4 (sim: 0.85) --}\\[1pt]
    \textcolor{keycolor}{Product:}\ \ \textcolor{valcolor}{Stainless steel insulated food jar, 300\,ml}\\
    \textcolor{keycolor}{Category:}\ \textcolor{valcolor}{Kitchenware}\\
    \textcolor{keycolor}{PCF:}\ \ \ \ \ \textcolor{excolor}{\textbf{5.0\,kg CO\textsubscript{2}e}}
  }\\[2pt]
  {\small\ttfamily%
    \textcolor{commentcolor}{-- Example 5 (sim: 0.82) --}\\[1pt]
    \textcolor{keycolor}{Product:}\ \ \textcolor{valcolor}{Plastic reusable travel mug with silicone sleeve, 355\,ml}\\
    \textcolor{keycolor}{Category:}\ \textcolor{valcolor}{Kitchenware}\\
    \textcolor{keycolor}{PCF:}\ \ \ \ \ \textcolor{excolor}{\textbf{1.4\,kg CO\textsubscript{2}e}}
  }
  \vspace{5pt}\\
  \promptrule\\
  \vspace{3pt}

  \sectionpill{querybg}{queryborder}{Query product}{querycolor}\\[4pt]
  {\small\ttfamily%
    \textcolor{keycolor}{Product:}\ \ \textcolor{querycolor}{\textbf{Stainless steel double-wall travel mug, 16\,oz (473\,ml), leak-proof lid}}\\
    \textcolor{keycolor}{Category:}\ \textcolor{querycolor}{\textbf{Kitchenware}}\\
    \textcolor{keycolor}{PCF:}\ \ \ \ \ \textcolor{querycolor}{\textbf{?}}
  }
  \vspace{5pt}\\
  \promptrule\\
  \vspace{3pt}

  \sectionpill{headerbg}{headerborder}{Response format}{rolecolor}\\[4pt]
  {\small\ttfamily%
    \textcolor{commentcolor}{Step-by-step reasoning:}\ \textcolor{valcolor}{[explain material, weight, and category reasoning]}\\
    \textcolor{commentcolor}{PCF:}\ \textcolor{valcolor}{X.X\,kg CO\textsubscript{2}e}
  }
  \vspace{8pt}

  \end{minipage}
  \hspace{12pt}%
};
\end{tikzpicture}}
  \caption{Few-shot LLM prompt template for PCF estimation.}
  \label{fig:llm-prompt-appendix}
\end{figure}

\end{appendices}

\end{document}